\documentclass[english]{article}
\usepackage[latin9]{inputenc}
\usepackage{geometry}
\usepackage{float}
\usepackage{mathrsfs}
\usepackage{amsmath}
\usepackage{amssymb}
\usepackage{graphicx}

\makeatletter

\floatstyle{ruled}
\newfloat{algorithm}{tbp}{loa}
\providecommand{\algorithmname}{Algorithm}
\floatname{algorithm}{\protect\algorithmname}

\@ifundefined{date}{}{\date{}}

\usepackage{float}\usepackage{mathrsfs}\usepackage{amsfonts}\usepackage{bm}\usepackage{subfigure}\usepackage{amsthm}
\usepackage{epsfig}

\@ifundefined{definecolor}{\@ifundefined{definecolor}
 {\@ifundefined{definecolor}
 {\usepackage{color}}{}
}{}
}{}

\usepackage[all]{xy}

\newtheorem{rem}{Remark}[section]
\newtheorem{prop}{Proposition}[section]

\newcounter{hypA}
\newenvironment{hypA}{\refstepcounter{hypA}\begin{itemize}
  \item[({\bf A\arabic{hypA}})]}{\end{itemize}}

\usepackage{babel}

\usepackage{babel}

\usepackage{babel}

\makeatother

\usepackage{babel}
\begin{document}
\begin{center}
\textbf{\Large Approximate Inference for Observation Driven Time Series
Models with Intractable Likelihoods}
\par\end{center}{\Large \par}

\begin{center}

\par\end{center}

\begin{center}
BY AJAY JASR$\textrm{A}^{1}$, NIKOLAS KANTA$\textrm{S}^{2}$, \&
ELENA EHRLIC$\textrm{H}^{3}$
\par\end{center}

\begin{center}
{\footnotesize $^{1}$Department of Statistics \& Applied Probability,
National University of Singapore, Singapore, 117546, SG.}\\
 {\footnotesize E-Mail:\,}\texttt{\emph{\footnotesize staja@nus.edu.sg}}
\par\end{center}{\footnotesize \par}

\begin{center}
{\footnotesize $^{2}$Department of Statistical Science, University
College London, London, WC1E 6BT, UK.}\\
 {\footnotesize E-Mail:\,}\texttt{\emph{\footnotesize n.kantas@ucl.ac.uk}}
\par\end{center}{\footnotesize \par}

\begin{center}
{\footnotesize $^{3}$Department of Mathematics, Imperial College
London, London, SW7 2AZ, UK.}\\
 {\footnotesize E-Mail:\,}\texttt{\emph{\footnotesize e.ehrlich05@imperial.ac.uk}}
\par\end{center}{\footnotesize \par}
\begin{abstract}
In the following article we consider approximate Bayesian parameter inference for observation driven time series models.
Such statistical models appear in a wide variety of applications, including econometrics and applied mathematics.
This article considers the scenario where the likelihood function cannot be evaluated point-wise; in such cases,
one cannot perform exact statistical inference, including parameter estimation, which often requires advanced computational algorithms, such as Markov chain Monte Carlo (MCMC). 
We introduce a new approximation based upon approximate Bayesian computation (ABC). Under some conditions, we show that 
as $n\rightarrow\infty$, with $n$ the length of the time series, the ABC posterior has, almost surely, a maximum \emph{a posteriori} (MAP) estimator of the parameters which is different from the true parameter.
 However, a noisy ABC MAP, which perturbs the original data, asymptotically converges to the true parameter, almost surely. In order to draw statistical inference,
for the ABC approximation adopted, standard MCMC algorithms can have acceptance probabilities that fall at an exponential rate in $n$
and slightly more advanced algorithms can mix poorly.
We develop a new and improved MCMC kernel, which is based upon an exact approximation of a marginal algorithm, whose cost per-iteration is random but the expected cost, for good performance, is
shown to be $\mathcal{O}(n^2)$ per-iteration. 
We  implement our new MCMC kernel for parameter inference from models in econometrics.
\end{abstract}

\section{Introduction}

\label{sec:intro}

Observation driven time-series models, introduced by \cite{cox}, has a wide variety of real applications, including econometrics (GARCH models) and applied mathematics (inferring
initial conditions and parameters of ordinary differential equations). The model can be described as follows. We observe $\{Y_k\}_{k\in\mathbb{N}_0}$, $Y_k\in\mathsf{Y}$ which are associated to a dynamic system $\{X_k\}_{k\in \mathbb{N}_0}$, $X_k\in\mathsf{X}$ which is potentially unknown.
Define the process $\{Y_k,X_k\}_{n\geq 0}$ (with $y_0$ some arbitrary point on $\mathsf{Y}$) on a probability space $(\Omega,\mathscr{F},\mathbb{P}_{\theta})$, where, for every $\theta\in\Theta\subseteq\mathbb{R}^{d_{\theta}}$, $\mathbb{P}_{\theta}$ is a probability measure. Denote by $\mathscr{F}_k = \sigma(\{Y_n,X_n\}_{0\leq n \leq k})$.
The model is defined as, for $k\in\mathbb{N}_0$
\begin{eqnarray*}
\mathbb{P}(Y_{k+1}\in A|\mathscr{F}_k) & = & \int_A H^{\theta}(x_k,dy) \quad A\times\mathsf{X}\in\mathscr{F}\\
X_{k+1} & = & \Phi^{\theta}(X_k,Y_{k+1}) \\
\mathbb{P}_{\theta}(X_0\in B) & = & \int_B \Pi_{\theta}(dx) \quad\mathsf{Y}\times B\in\mathscr{F}
\end{eqnarray*}
where $H:\Theta\times\mathsf{X}\times\sigma(\mathsf{Y})\rightarrow[0,1]$, $\Phi:\Theta\times\mathsf{X}\times\mathsf{Y}\rightarrow\mathsf{X}$ and for every $\theta\in\Theta$, $\Pi_{\theta}\in\mathcal{P}(\mathsf{X})$ (the probabilities on $\mathsf{X}$). Throughout, we assume that
for any $(x,\theta)\in\mathsf{X}\times\Theta$ $H^{\theta}(x,\cdot)$ admits a density w.r.t.~some $\sigma-$finite measure $\mu$, which we denote as $h^{\theta}(x,y)$.
Next, we define a prior probability distribution $\Xi$ on $(\Theta,\mathcal{B}(\Theta))$, with Lebesgue density $\xi$. Thus, given $n$ observations $y_{1:n}:=(y_1,\dots,y_n)$ the object of inference is the posterior distribution on $\Theta\times \mathsf{X}$:
\begin{equation}
\Pi(d(\theta,x_0)|y_{1:n}) \propto \Bigg(\prod_{k=1}^n h^{\theta}(\Phi^{\theta}(y_{0:k-1})(x_0),y_k)\Bigg)\Pi_{\theta}(dx_0)\xi(\theta) d\theta
\label{eq:true_post}
\end{equation}
where we have used the notation $\Phi^{\theta}(y_{0:k-1})(x_0) = \Phi^\theta \circ\cdots\circ \Phi^{\theta}(x_0,y_1)$ and $d\theta$ is Lebesgue measure. In most applications of practical interest, one cannot compute the posterior point-wise and has to resort
to numerical methods, such as MCMC, to draw inference on $\theta$ and/or $x_0$.

In this article, we are not only interested in inferring the posterior distribution, but the scenario for which $h^{\theta}(x,y)$ cannot be evaluated point-wise, nor do we have access to an unbiased estimate of it (it is assumed we can simulate from the associated distribution).
In such a case, it is not possible to draw inference from the true posterior, even using numerical techniques. The common response in Bayesian statistics, is now to adopt an approximation of the posterior
using the notion of approximate Bayesian computation (ABC); see \cite{marin} for a recent overview. ABC approximations of posteriors are based upon defining a probability distribution on an extended state-space, with
the additional random variables lying on the data-space and usually distributed according the true likelihood. The closeness of the ABC posterior distribution is controlled by a tolerance parameter $\epsilon>0$ and often
the approximation is exact as $\epsilon\rightarrow 0$.

In this paper, we introduce a new ABC approximation of observation driven time-series models, which is closely associated to that developed in \cite{jasra} for hidden Markov models (HMMs) and later for static parameter inference from HMMs \cite{dean}. This latter ABC
approximation is particularly well behaved and a noisy variant (which pertrubs the data; see e.g.~\cite{dean}) is shown under some assumptions to provide maximum-likelihood estimators (MLE) which asympotically in $n$ are the true parameters. The new ABC approximation that we develop is studied from a theoretical perspective.
Relying on the recent work of \cite{douc} we show that, under some conditions, 
as $n\rightarrow\infty$, with $n$ the length of the time series, the ABC posterior has, almost surely, a MAP estimator of $\theta$ which is different from the true parameter $\theta^*$ say.
However, a noisy ABC MAP of $\theta$ asymptotically converges to the true parameter, almost surely. These results establish that the particular approximation adopted is reasonably sensible.

The other main contribution of this article is a development of a new MCMC algorithm designed to sample from the ABC approximation of the posterior. Due to the nature of the ABC approximation it is easily seen that standard MCMC algorithms (e.g.~\cite{majoram}) will have an acceptance probability that will
fall at an exponential rate in $n$. In addition, more advanced ideas such as those based upon the `pseudo marginal' \cite{beaumont}, have recently been shown to perform rather poorly in theory; see \cite{lee1}. These latter algorithms are based upon exact approximations of marginal algorithms \cite{andrieu,andrieu1},
which in our context is just sampling $\theta,x_0$. We develop an MCMC kernel, related to recent work in \cite{lee}, which is designed to have a random running time per-iteration, with the idea of improving the exploration ability of the Markov chain. We show that the expected cost per iteration of the algorithm,
under some assumptions and for reasonable performance, is $\mathcal{O}(n^2)$, which compares favourably with competing algorithms.
We also show, empirically, that this new MCMC method out-performs standard pseudo marginal algorithms.

This paper is structured as follows. In Section \ref{sec:model} we introduce our ABC approximation and give our theoretical results on the MAP estimator. In Section \ref{sec:comp}, we give our new MCMC algorithm, along with some theoretical discussion about its computational cost and stability.
In Section \ref{sec:examples} our approximation and MCMC algorithm is illustrated on toy and real examples. In Section \ref{sec:summ} we conclude the article with some discussion of future work. The proofs of our theoretical results are given in the appendix.

\section{Approximate posteria using ABC approximations}

\label{sec:model}

\subsection{ABC approximations and noisy ABC}

As it was emphasised in Section \ref{sec:intro}, we are interested
in performing inference when $h^{\theta}(x,y)$ cannot be evaluated
point-wise, nor do we have access to an unbiased estimate of it. We
will instead assume it is possible to sample from $h^{\theta}$. In
such scenaria, one cannot use standard simulation based methods. For
example, in a standard MCMC approach the Metropolis-Hastings acceptance
ratio cannot be evaluated, even though it may be well-defined. Following
the work in \cite{dean,jasra} for hidden Markov models,
we introduce an ABC approximation for the density of the posterior in \eqref{eq:true_post} as follows:  
\begin{equation}
\pi_{n}^{\epsilon}(\theta,x_{0}|y_{1:n})\propto\prod_{k=1}^{n}h^{\theta,\epsilon}\left(\Phi^{\theta}\left(y_{0:k-1}\right)(x_{0}),y_{k}\right)\xi(x_{0},\theta),\label{eq:abc_post1}
\end{equation}
 with $\epsilon>0$ and 
\begin{equation}
h^{\theta,\epsilon}(\Phi^{\theta}(y_{0:k-1})(x_{0}),y_{k})=\frac{\int_{B_{\epsilon}(y_{k})}h^{\theta}(\Phi^{\theta}(y_{0:k-1})(x_{0}),y)\mu(dy)}{\mu(B_{\epsilon}(0))},\label{eq:pseudo_like_def}
\end{equation}
where we denote $B_{\epsilon}(y)$ as the open ball centred at $y$
with radius $\epsilon$ and write $\mu(B_{\epsilon}(y))=\int_{B_{\epsilon}(y)}\mu(dx)$.
When $\mu$ is the Lebesgue measure, $\mu(B_{\epsilon}(y))$ corresponds
to the volume of the ball $B_{\epsilon}(y)$. 

In general we will refer to ABC as the procedure of performing inference
for the posterior in \eqref{eq:abc_post1}. In addition, we will call
\emph{noisy ABC} the inference procedure that uses instead of the
original observation sequence a perturbed one, namely $\left\{ \hat{Y}_{k}\right\} _{k\geq0}$,
where each $\hat{Y}_{k}$ is given by 
\[
\hat{Y}_{k}=Y_{k}+\epsilon Z_{k},
\]
with each $Z_{k}$ is identically independently distributed (i.i.d.)
uniformly on $B_{\epsilon}(0)$ (shorthand $Z_{k}\sim\mathcal{U}_{B_{\epsilon}(0)}$).

\subsection{Consistency results for the MAP estimator}

\label{sec:consis}

In this section we will investigate some interesting properties of
the ABC posterior in \eqref{eq:abc_post1}. In particular, we will
look at the  asymptotic behaviour with $n$ of the resulting MAP estimators
for $\theta$. The properties of the MAP estimator reveal information
about the mode of the posterior distribution as we obtain increasingly
more data. To simplify the analysis in this section we will assume
that:

\begin{hypA}\label{as:1-1}
\begin{itemize}
\item $x_{0}$ is fixed and known, i.e. $\Pi(dx_{0})=\delta_{x}(dx_{0})$,
where $\delta_{x}$ denotes the Dirac delta measure on $\mathsf{X}$
and $x\in\mathsf{X}$ is known.  
\item $\xi(x,\cdot)$ is bounded and positive everywhere in $\Theta$.
\item the observations actually originated from the true model model for
some $\theta^{*}\in\Theta$, i.e. we look at a well-specified problem.
\item $H$ and $h$ do not depend upon $\theta$. Thus we have the following
model recursions for the true model: 
\begin{eqnarray}
\mathbb{P}_{\theta^{*}}(Y_{k+1}\in A|\mathscr{F}_{k}) & = & \int_{A}H(x_{k},dy),\quad A\times\mathsf{X}\in\mathscr{F},\nonumber \\
X_{k+1} & = & \Phi^{\theta^{*}}(X_{k},Y_{k+1}),\label{eq:mod_rec}
\end{eqnarray}
where we will denote associated expectations to $\mathbb{P}_{\theta^{*}}$
as $\mathbb{E}_{\theta^{*}}$.
\end{itemize}
\end{hypA}

In addition, for this section we will introduce some extra notations:
$(\mathsf{X},d)$ is a compact, complete and separable metric space
and $(\Theta,\mathsf{d})$ is a compact metric space, with $\Theta\subset\mathbb{R}^{d_{\theta}}$.
For two measures $\nu$ and $\lambda$ of bounded variation denote
the convolution $\nu\star\lambda\left(f\right)=\int f(z+r)\nu(dz)\lambda(dr)$.
Let also $\mathbb{Q}_{\epsilon}$ be the probability law associated
to the random sequence $\{Z_{k}\}_{k\in\mathbb{Z}}$ , where each
$Z_{k}$ is an i.i.d. sample from the uniform distribution defined
on $B_{\epsilon}(0)$. 

We proceed with some additional technical assumptions:

\begin{hypA}\label{as:1} $\{X_{k},Y_{k}\}_{k\in\mathbb{Z}}$ is
a stationary stochastic process, with $\{Y_{k}\}_{k\in\mathbb{Z}}$
strict sense stationary and ergodic, following \eqref{eq:mod_rec}.
\end{hypA} 

\begin{hypA}\label{as:2} For every $(x,y)\in\mathsf{X}\times\mathsf{Y}$,
$\theta\mapsto\Phi^{\theta}(x,y)$ is continuous. In addition, there
exist $0<C<\infty$ such that for any $(x,x')\in\mathsf{X}$, $\sup_{y\in\mathsf{Y}}|h(x,y)-h(x',y)|\leq Cd(x,x')$.
Finally $0<\underline{h}\leq h(x,y)\leq\overline{h}<\infty$, for
every $(x,y)\in\mathsf{X}\times\mathsf{Y}$. \end{hypA}

\begin{hypA}\label{as:3} There exist a measurable $\varrho:\mathsf{Y}\rightarrow(0,1)$,
such that for every $(\theta,y,x,x')\in\Theta\times\mathsf{Y}\times\mathsf{X}^{2}$
\[
d(\Phi^{\theta}(x,y),\Phi^{\theta}(x',y))\leq\varrho(y)d(x'x').
\]
 \end{hypA}

\begin{hypA}\label{as:4} The following statements hold:
\begin{enumerate}
\item {$H(x,\cdot)=H(x',\cdot)$ if and only if $x=x'$} 
\item {If $\Phi^{\theta}(Y_{-\infty:0})(x)=\Phi^{\theta'}(Y_{-\infty:0})(x)$
holds $\mathbb{P}_{\theta^{*}}\star\mathbb{Q}_{\epsilon}-$a.s., then
$\theta=\theta^{'}$.} 
\end{enumerate}
\end{hypA} Assumptions (A\ref{as:1}-\ref{as:4}) and the compactness
of $\Theta$ are standard assumptions for maximum likelihood estimation
(ML) and they can be used to show the uniqueness of the maximum likelihood
estimator (MLE); see \cite{douc} for more details.
Therefore, if the prior $\xi(\theta)$ is bounded and positive everywhere
on $\Theta$ it is a simple corollary that the MAP estimator will
correspond to the MLE. In the remaining part of this section we will
adapt the analysis in \cite{douc} for MLE to the ABC setup. 

In particular, we are to estimate $\theta$ using the log-likelihood
function: 
\[
l_{\theta,x}(y_{1:n}):=\frac{1}{n}\sum_{k=1}^{n}\log\Big(h^{\epsilon}(\Phi^{\theta}(y_{0:k-1})(x),y_{k})\Big)
\]
We define the ABC-MLE for an $n$-long sequence as 
\[
\theta_{n,x,\epsilon}=\textrm{\emph{\ensuremath{\arg\max}}}_{\theta\in\Theta}l_{\theta,x}(y_{1:n}).
\]
We proceed with the following proposition:

\begin{prop}\label{prop1} Assume (A\ref{as:1-1}-\ref{as:3}). Then
for every $x\in\mathsf{X}$ and fixed $\epsilon>0$ 
\[
\lim_{n\rightarrow\infty}\mathsf{d}(\theta_{n,x,\epsilon},\Theta_{\epsilon}^{*})=0\quad\mathbb{P}_{\theta*}-a.s.
\]
 where $\Theta_{\epsilon}=\textrm{\emph{\ensuremath{\arg\max}}}_{\theta\in\Theta}\mathbb{E}_{\theta^{*}}[\log(h^{\epsilon}(\Phi^{\theta}(Y_{-\infty:0})(x),Y_{1}))]$. 

\end{prop}The result establishes that the estimate will converge
to a point, which is typically different to the true parameter. Hence
there is an intrinsic asymptotic bias for the plain ABC procedure.
To correct this bias, consider the noisy ABC procedure, of replacing
the observations by $\hat{Y}_{k}=Y_{k}+\epsilon Z_{k},$ where $Z_{k}\stackrel{\textrm{i.i.d.}}{\sim}\mathcal{U}_{B_{1}(0)}$.
 The noisy ABC MLE estimator is then: 
\[
\hat{\theta}_{n,x,\epsilon}=\textrm{\emph{\ensuremath{\arg\max}}}_{\theta\in\Theta}\frac{1}{n}\sum_{k=1}^{n}\log\Big(h^{\epsilon}(\Phi^{\theta}(\hat{y}_{0:k-1})(x),\hat{y}_{k})\Big).
\]
We have the following result:

\begin{prop}\label{prop2} Assume (A\ref{as:1-1}-\ref{as:4}). Then
for every $x\in\mathsf{X}$ and fixed $\epsilon>0$ 
\[
\lim_{n\rightarrow\infty}\mathsf{d}(\hat{\theta}_{n,x,\epsilon},\theta^{*})=0\quad\mathbb{P}_{\theta*}\star\mathbb{Q}_{\epsilon}-a.s..
\]
 \end{prop}

The result shows that the noisy ABC MLE estimator is asymptotically
unbiased. Therefore, given that in our setup the ABC MAP estimator
corresponds to the ABC MLE we can conclude that the mode of the posterior
distribution as we obtain increasingly more data is converging towards
the true parameter. Finally we note that our assumptions indeed pose
some restrictions, but these are shown to be realistic for a few interesting
models in \cite{douc}. In addition, the main purpose of this result is to motivate  the use of the approximate posterior
in \eqref{eq:abc_post1} when the observation sequence is long or
its marginal likelihood is quite informative.

\section{Computational Methodology}

\label{sec:comp}

Recall that we formulated in the ABC posterior written in \eqref{eq:abc_approx}.
One can rewrite the approximate posterior in \eqref{eq:abc_post1}:
\[
\pi^{\epsilon}(\theta,x_{0}|y_{1:n})=\frac{p_{\theta,x_{0}}^{\epsilon}(y_{1:n})\xi(x_{0},\theta)}{\int p_{\theta,x_{0}}^{\epsilon}(y_{1:n})\xi(x_{0},\theta)dx_{0}d\theta},
\]
with 
\[
p_{\theta,x_{0}}^{\epsilon}(y_{1:n})=\int\prod_{k=1}^{n}\frac{\mathbb{I}_{B_{\epsilon}(y_{k})}(u_{k})}{\mu(B_{\epsilon}(0))}h^{\theta}(\Phi^{\theta}(y_{0:k-1})(x_{0}),u_{k})du_{1:n}.
\]
Note we have just used Fubini's theorem to rewritte the likehood $p_{\theta,x_{0}}^{\epsilon}(y_{1:n})$
as an integral of a product instead of a product of integrals $\prod_{k=1}^{n}h^{\theta,\epsilon}\left(\Phi^{\theta}\left(y_{0:k-1}\right)(x_{0}),y_{k}\right)$
shown in \eqref{eq:abc_post1}-\eqref{eq:pseudo_like_def}. In this
paper we will focus only on MCMC algorithms and in particular on the
Metropolis-Hastings (M-H) approach. In order to sample from the posterior
$\pi^{\epsilon}$ one runs an ergodic Markov Chain with the invariant
density being $\pi^{\epsilon}$. Then after a few iterations when
the chain has reached stationarity, one can treat the samples from
the chain as approximate samples from $\pi^{\epsilon}$. This is shown
in Algorithm \ref{alg:mcmc}, where for convenience we denote $\gamma=(\theta,x_{0})$.
The one-step transition kernel of the MCMC chain is usually described
as the \emph{M-H kernel} and follows from Step 2 in Algorithm \ref{alg:mcmc}.

\begin{algorithm}
\begin{enumerate}
\item \textbf{(Initialisation)} At $t=0$ sample $\gamma_{0}\sim\xi$.
\item \textbf{(M-H kernel)} For $t\geq1$:

\begin{itemize}
\item Sample $\gamma'|\gamma_{t-1}$ from a proposal $Q(\gamma_{t-1},\cdot)$ with density
$q(\gamma_{t-1},\cdot)$.
\item Accept the proposed state and set $\gamma_{t}=\gamma'$ with probability
\[
1\wedge\frac{p_{\gamma'}^{\epsilon}(y_{1:n})}{p_{\gamma_{t-1}}^{\epsilon}(y_{1:n})}\times\frac{\xi(\gamma')q(\gamma',\gamma_{t-1})}{\xi(\gamma_{t-1})q(\gamma_{t-1},\gamma')},
\]
otherwise set $\gamma_{t}=\gamma_{t-1}$. Set $t=t+1$ and return to the start of 2.
\end{itemize}
\end{enumerate}
\caption{A marginal M-H algorithm for $\pi^{\epsilon}(\gamma|y_{1:n})$}
\label{alg:mcmc}

\end{algorithm}

Unfortunately $p_{\theta,x_{0}}^{\epsilon}(y_{1:n})$ is not available
analytically and cannot be evaluated, so this rules out the possibility
of using of traditional MCMC approaches like Algorithm \ref{alg:mcmc}.
However, one can resort to the so called \emph{pseudo-marginal} approach
whereby unbiased estimates of $p_{\theta,x_{0}}^{\epsilon}(y_{1:n})$
are used instead within an MCMC algorithm. We will refer to this algorithm
as ABC-MCMC. The resulting algorithm can be posed as one targeting
a posterior defined on an extended state space, so that its marginal
coincides with $\pi^{\epsilon}(\theta,x_{0}|y_{1:n})$. We will use
these ideas to present ABC-MCMC as a M-H algorithm which is an exact
approximation to an appropriate marginal algorithm. 

To illustrate an example of these ideas, we proceed by writing a posterior
on an extended state-space $\Theta\times\mathsf{X}\times\mathsf{Y}^{n}$
as follows: 
\begin{equation}
\pi_{n}^{\epsilon}(\theta,x_{0},u_{1:n}|y_{1:n})\propto\prod_{k=1}^{n}\mathbb{I}_{B_{\epsilon}(y_{k})}(u_{k})h^{\theta}\left(\Phi^{\theta}\left(y_{0:k-1}\right)(x_{0}),y_{k}\right)\xi(x_{0},\theta).\label{eq:abc_approx}
\end{equation}
It is clear that \eqref{eq:abc_post1} is the marginal of \eqref{eq:abc_approx}
and hence the similarity in the notation. As we will show later in
this section, extending the target space in the posterior as in \eqref{eq:abc_approx}
is not the only and certainly not the best choice. We emphasise that
the only essential requirement for each choice is that the marginal
of the extended target is $\pi^{\epsilon}(\theta,x_{0}|y_{1:n})$,
but one should be cautious because the particular choice will affect
the mixing properties and the efficiency of the MCMC scheme that will
be used to sample from $\pi_{n}^{\epsilon}(\theta,x_{0},u_{1:n}|y_{1:n})$
in \eqref{eq:abc_approx} or another variant.

\subsection{Standard approaches for ABC-MCMC}

\label{sec:mcmc}

We will now look at two basic different choices for extending the
ABC posterior while keeping the marginal fixed to $\pi^{\epsilon}(\theta,x_{0}|y_{1:n})$.
In the remainder of the paper we will denote $\gamma=(\theta,x_{0})$
as we did in Algorithm \ref{alg:mcmc}.

Initially consider the ABC approximation when be extended to the space
$\Theta\times\mathsf{X}\times\mathsf{Y}^{n}$: 
\[
\pi^{\epsilon}(\gamma,u_{1:n}|y_{1:n})=\frac{\xi(\gamma)p_{\gamma}^{\epsilon}(y_{1:n})}{\int\xi(\gamma)p_{\gamma}^{\epsilon}(y_{1:n})d\gamma}\frac{\prod_{k=1}^{n}\frac{\mathbb{I}_{B_{\epsilon}(y_{k})}(u_{k})}{\mu(B_{\epsilon}(0))}}{p_{\gamma}^{\epsilon}(y_{1:n})}\prod_{k=1}^{n}h^{\theta}(\Phi^{\theta}(y_{0:k-1})(x_{0}),u_{k}).
\]
Recall one cannot evaluate $h^{\theta}(\Phi^{\theta}(x_{0})(u_{k}),u_{k})$
and is only able to simulate from it. In Algorithm \eqref{alg:simple}
we present a natural M-H proposal that could be used to sample from
$\pi^{\epsilon}(\gamma,u_{1:n}|y_{1:n})$ instead of the one shown
Step 2 at Algorithm \ref{alg:mcmc}. Note that this time the state
of the MCMC chain is composed of $\left(\gamma,u_{1:n}\right)$. Here
each $u_{k}$ assumes the role of an auxiliary variable to be eventually
integrated out at the end of the MCMC procedure. 

\begin{algorithm}
\begin{itemize}
\item {Sample $\gamma'|\gamma$ from a proposal $Q(\gamma,\cdot)$ with
density $q(\gamma,\cdot)$.} 
\item {Sample $u_{1:n}^{'}$ from a distribution with joint density $\prod_{k=1}^{n}h^{\theta'}(\Phi^{\theta'}(y_{0:k-1})(x_{0}),u_{k})$
} 
\item {Accept the proposed state $\left(\gamma',u'_{1:n}\right)$ with
probability: 
\[
1\wedge\frac{\prod_{k=1}^{n}\mathbb{I}_{B_{\epsilon}(y_{k})}(u_{k}^{'})}{\prod_{k=1}^{n}\mathbb{I}_{B_{\epsilon}(y_{k})}(u_{k})}\times\frac{\xi(\gamma')q(\gamma',\gamma)}{\xi(\gamma)q(\gamma,\gamma')}.
\]
 } 
\end{itemize}
\caption{M-H Proposal for basic ABC MCMC}
\label{alg:simple}
\end{algorithm}
However, as $n$ increases, the M-H kernel in Algorithm \ref{alg:simple}
will have an acceptance probability that falls quickly with $n$.
In particular, for any fixed $\gamma$, the probability of obtaining
such a sample will fall at an exponential rate in $n$. This means
that this basic ABC MCMC approach will be inefficient for a moderate
value of $n$.

 This issue can be dealt with by using $N$ multiple trials, so that
at each $k$, some auxiliary variables (or pseudo-observations) are
in the ball $B_{\epsilon}(y_{k})$. This idea originates from \cite{beaumont,majoram}
and in fact augments the posterior to a larger state-space, $\Theta\times\mathsf{X}\times\mathsf{Y}^{nN}$,
in order to target   the following density: 
\[
\widetilde{\pi}^{\epsilon}(\gamma,u_{1:n}^{1:N}|y_{1:n})=\frac{\pi(\gamma)p_{\gamma}^{\epsilon}(y_{1:n})}{\int\pi(\gamma)p_{\gamma}^{\epsilon}(y_{1:n})d\gamma}\frac{\prod_{k=1}^{n}\frac{\sum_{j=1}^{N}\mathbb{I}_{B_{\epsilon}(y_{k})}(u_{k}^{j})}{N\mu(B_{\epsilon}(0))}}{p_{\gamma}^{\epsilon}(y_{1:n})}\prod_{k=1}^{n}\prod_{j=1}^{N}h^{\theta}(\Phi^{\theta}(y_{0:k-1})(x_{0}),u_{k}^{'j})
\]
Again, it is easy to show that the marginal of interest $\pi^{\epsilon}(\gamma|y_{1:n})$
is preserved, i.e. 
\[
\pi^{\epsilon}(\gamma|y_{1:n})=\int_{\mathsf{Y}^{nN}}\widetilde{\pi}^{\epsilon}(\gamma,u_{1:n}^{1:N}|y_{1:n})du_{1:n}^{1:N}=\int_{\mathsf{Y}^{n}}\pi^{\epsilon}(\gamma,u_{1:n}|y_{1:n})du_{1:n}.
\]
In Algorithm \ref{alg:Ntry} we present an M-H kernel with invariant
density $\widetilde{\pi}^{\epsilon}$. The state of the MCMC chain
now is $\left(\gamma,u_{1:n}^{1:N}\right)$. We remark that as $N$
grows, one expects to recover the properties of the ideal M-H algorithm
in Algorithm \ref{alg:mcmc}. Nevertheless, it has been shown in \cite{lee1}
that even the M-H kernel in Algorithm \ref{alg:Ntry} does not always
perform well. It can happen that the chain gets often stuck in regions
of the state-space $\Theta\times\mathsf{X}$ where 
\[
\alpha_{k}(y_{1:k},\epsilon,\gamma):=\int_{B_{\epsilon}(y_{k})}h^{\theta}(\Phi^{\theta}(y_{0:k-1})(x_{0}),u)du
\]
is small. 
\begin{algorithm}
\begin{itemize}
\item {Sample $\gamma'|\gamma$ from a proposal $Q(\gamma,\cdot)$ with
density $q(\gamma,\cdot)$.} 
\item {Sample $\left.u'\right._{1:n}^{1:N}$ from a distribution with joint
density $\prod_{k=1}^{n}\prod_{j=1}^{N}h^{\theta'}(\Phi^{\theta'}(y_{0:k-1})(x_{0}),\left.u'\right._{k}^{j})$
.} 
\item {Accept the proposed state $\left(\gamma',\left.u'\right._{1:n}^{1:N}\right)$
with probability: 
\[
1\wedge\frac{\prod_{k=1}^{n}(\frac{1}{N}\sum_{j=1}^{N}\mathbb{I}_{B_{\epsilon}(y_{k})}(\left.u'\right._{k}^{j}))}{\prod_{k=1}^{n}(\frac{1}{N}\sum_{j=1}^{N}\mathbb{I}_{B_{\epsilon}(y_{k})}(u_{k}^{j}))}\times\frac{\pi(\gamma')q(\gamma',\gamma)}{\pi(\gamma)q(\gamma,\gamma')}.
\]
 } 
\end{itemize}
\caption{M-H Proposal for ABC with N trials}
\label{alg:Ntry}
\end{algorithm}

\subsection{A Metropolis-Hastings kernel for ABC with a random number of trials}

\label{sec:new_kernel}

We will address this shortfall detailed above, by proposing an alternative augmented
target and corresponding M-H kernel. The basic idea is that a random
number of trials is used based on the value of $\alpha_{k}(y_{1:k},\epsilon,\gamma)$.
Then it will be possible to use more computational effort when the
chain is at regions where $\alpha_{k}(y_{1:k},\epsilon,\gamma)$ is
low.

Consider an alternative extended target, for $N\geq2$, $m_{k}\in\mathsf{M}_{N}:=\{N,N+1,\dots,\}$,
$1\leq k\leq n$: 
\[
\hat{\pi}^{\epsilon}(\gamma,m_{1:n}|y_{1:n})=\frac{\pi(\gamma)p_{\gamma}^{\epsilon}(y_{1:n})}{\int\pi(\gamma)p_{\gamma}^{\epsilon}(y_{1:n})d\gamma}\frac{\prod_{k=1}^{n}\frac{N-1}{\mu(B_{\epsilon}(0))(m_{k}-1)}}{p_{\gamma}^{\epsilon}(y_{1:n})}\prod_{k=1}^{n}\binom{m_{k}-1}{N-1}\alpha_{k}(y_{1:k},\gamma,\epsilon)^{N}(1-\alpha_{k}(y_{1:k},\gamma,\epsilon))^{m_{k}-N}.
\]
Standard results for negative binomial distrubutions (see \cite{neuts,zacks}
for more details) imply that 
\begin{equation}
\sum_{m_{k}=N}^{\infty}\frac{1}{m_{k}-1}\binom{m_{k}-1}{N-1}\alpha_{k}(y_{1:k},\epsilon,\gamma)^{N}(1-\alpha_{k}(y_{1:k},\epsilon,\gamma))^{m_{k}-N}=\frac{\alpha_{k}(y_{1:k},\epsilon,\gamma)}{N-1}\label{eq:neg_binomial_recursion}
\end{equation}
holds and this can be used to deduce that  
\[
\prod_{k=1}^{n}\frac{N-1}{\mu(B_{\epsilon}(0))(m_{k}-1)}
\]
is an unbiased estimator for $p_{\gamma}^{\epsilon}(y_{1:n})$. In
addition, from \eqref{eq:neg_binomial_recursion} it follows that
the marginal w.r.t. $\gamma$ is the one of interest: 
\[
\pi^{\epsilon}(\gamma|y_{1:n})=\sum_{m_{1:n}\in\mathsf{M}_{N}^{n}}\widehat{\pi}^{\epsilon}(\gamma,m_{1:n}|y_{1:n})
\]
In Algorithm \ref{alg:new} we present a M-H kernel with invariant
density $\widehat{\pi}^{\epsilon}$. The state of the MCMC chain this
time is $\left(\gamma,m_{1:n}\right)$. 
\begin{algorithm}
\begin{itemize}
\item {Sample $\gamma'|\gamma$ from a proposal $Q(\gamma,\cdot)$ with
density $q(\gamma,\cdot)$.} 
\item {For $k=1,\dots,n$ repeat the following: sample $u_{k}^{1},u_{k}^{2},\dots$
with probability density $h^{\theta'}(\Phi^{\theta'}(y_{0:k-1})(x_{0}'),u_{k})$
until there are $N$ samples lying in $B_{\epsilon}(y_{k})$; the
number of samples to achieve this (including the successful trial)
is $m_{k}'$. } 
\item {Accept $\left(\gamma',m_{1:n}'\right)$ with probability: 
\[
1\wedge\frac{\prod_{k=1}^{n}\frac{1}{m_{k}'-1}}{\prod_{k=1}^{n}\frac{1}{m_{k}-1}}\times\frac{\pi(\gamma')q(\gamma',\gamma)}{\pi(\gamma)q(\gamma,\gamma')}.
\]
 } 
\end{itemize}
\caption{M-H Proposal with a random number of trials}
\label{alg:new}
\end{algorithm}

The potential benefit of this kernel is that one expects the probability
of accepting a proposal is higher than the previous M-H kernel (for
a given $N$). This comes at a computational cost which is both increased
and random. The proposed kernel is based on the \emph{$N-$hit} kernel
of \cite{lee}, which   has been adapted here to account for the data
being a sequence of observations resulting from a time series. 
 Finally, it is important to mention that in Algorithms \ref{alg:Ntry}
and \ref{alg:new} generating multiple trials can be implemented very
efficiently in parallel using appropriate computing hardware, such
as multi-core processors or computing clusters.

\subsubsection{On the choice of $N$}

To implement the proposed kernel, one needs to select $N$. In practice
to it is difficult to know a good value this \emph{a priori}, so we
present a theoretical result that can add some intuition on choosing
$N$. Let $\mathbb{E}_{\gamma,N}[\cdot]$ denote expectation w.r.t.~$\prod_{k=1}^{n}\binom{m_{k}-1}{N-1}\alpha_{k}(y_{1:k},\epsilon,\gamma)^{N}(1-\alpha_{k}(y_{1:k},\epsilon,\gamma))^{m_{k}-N}$
given $\gamma,N$. We will also pose the assumption: \begin{hypA}\label{ass:a_k_positive}
For any fixed $\epsilon>0$, $\gamma\in\Theta\times\mathsf{X}$, we
have $\alpha_{k}(y_{1:k},\epsilon,\gamma)>0$. \end{hypA} The the
following result holds, whose proof can be found in the appendix.: 

\begin{prop}\label{prop:rel_var} Assume (A\ref{ass:a_k_positive})
and let $\beta\in(0,1)$, $n\geq1$ and $N\geq\frac{2n}{1-\beta}\vee3$.
Then for fixed $(\gamma,\epsilon)\in\Theta\times\mathsf{X}\times\mathbb{R}^{+}$
we have  
\[
\mathbb{E}_{\gamma,N}\bigg[\bigg(\frac{\prod_{k=1}^{n}\frac{1}{M_{k}-1}}{\prod_{k=1}^{n}\frac{\alpha_{k}(y_{1:k},\epsilon,\gamma)}{N-1}}-1\bigg)^{2}\bigg]\leq\frac{Cn}{N}
\]
 where $C=1/\beta$. \end{prop}

The result shows that one should set $N=\mathcal{O}(n)$ for the relative
variance not to grow with $n$, which is unsuprising, given the conditional
independence structure of the $m_{1:n}$.
To get a better handle on the variance, suppose $n=1$, then for $\gamma$ fixed
$$
\mathbb{V}\textrm{\emph{ar}}_{\gamma,N}\Big[\frac{1}{N}\sum_{j=1}^N\mathbb{I}_{B_{\epsilon}(y_1)}(u_1^j)\Big]
= \frac{\alpha_1(y_{1},\epsilon,\gamma)(1-\alpha_1(y_{1},\epsilon,\gamma))}{N}.
$$
For the new approach one can show
$$
\mathbb{V}\textrm{\emph{ar}}_{\gamma,N}\Big[\frac{N-1}{M_1-1}\Big] \leq \frac{\alpha_1(y_{1},\epsilon,\gamma)^2}{(N-2)}.
$$
Not taking into account the computational cost, one prefers this new estimate with regards to variance if
$$
\frac{N}{N-2} \leq \frac{1-\alpha_1(y_{1},\epsilon,\xi)}{\alpha_1(y_{1},\epsilon,\gamma)}
$$
which is likely to occur if $\alpha_1(y_{1},\epsilon,\gamma)$ is not too large (recall we want $\epsilon$ to be small, so that we have a good approximation of the true posterior) and $N$ is moderate - this is precisely the scenario in practice.

%

\begin{rem}\label{rem:gamma_indep} It is easily shown that the relative
variance associated to the estimate $\prod_{k=1}^{n}\bigg[\Big(\frac{1}{N}\sum_{j=1}^{N}\mathbb{I}_{B_{\epsilon}(y_{k})}(u_{k}^{j})\Big)\bigg]$
is 
\[
\prod_{k=1}^{n}\Big[\frac{1}{\alpha_{k}(y_{1:k},\epsilon,\gamma)N}-\frac{N-1}{N}\Big]-1.
\]
Note this quantity is not uniformly upper-bounded in $\gamma$ unless
$\inf_{k,\gamma}\alpha_{k}(y_{1:k},\epsilon,\gamma)\geq C>0$, which
may not occur. Conversely, Proposition \ref{prop:rel_var} shows that
the relative variance of the new estimator is uniformly upper-bounded
in $\gamma$ under minimal conditions. We suspect that this means
in practice that the kernel with random number of trials may mix faster
. 
\end{rem}

\subsubsection{Computational considerations}\label{sec:comp_consider}

As the cost per-iteration is random, we will investigate this further.
We denote the proposal of $\gamma,m_{1:n}$ as $\tilde{Q}$. Let $\zeta$
be the initial distribution of the MCMC chain and $\zeta K^{t}$ the
distribution of the state at time $t$. In addition, denote by $m_{k}^{t}$
the proposed state for $m_{k}$ at iteration $t$. Finally, we will
write as $\mathbb{E}_{\zeta K^{t}\otimes\tilde{Q}}$ the expectation
of a random variable proposed by $\tilde{Q}$ given the simulated
state at time $t$. We will assume that the observations are fixed
and known. Then we have the following result:

\begin{prop}\label{prop:time} Let $\epsilon>0$, and suppose that
there exists a constant $C>0$ such that for any $n\geq1$ we have
$\inf_{k}\alpha_{k}(y_{1:k},\gamma,\epsilon)\geq C$, $\mu-$a.e..
Then it holds for any $N\geq2$, $t\geq1$, that: 
\[
\mathbb{E}_{\zeta K^{t}\otimes\tilde{Q}}[\sum_{k=1}^{n}M_{k}^{t}]\leq\frac{nN}{C}.
\]
 \end{prop}

The expected computational cost grows linearly with $n$. Thus, coupled with the result in Proposition \ref{prop:rel_var}, one has a cost of $\mathcal{O}(n^2)$ per-iteration,
which is comparable to many exact approximations of MCMC algorithms (e.g.~\cite{andrieu}), albeit in a much simpler situation. 
Note also
that the kernel in Algorithm \ref{alg:Ntry} is expected to require
a cost of $\mathcal{O}(n^{2})$ per iteration for reasonable performance,
although this cost here is deterministic. As mentioned above, one
expects the approach with random number of trials to work better 
with regards to the mixing time, especially when the values of $\alpha_{k}(y_{1:k},\epsilon,\gamma)$
are not large. We attribute this to Algorithm \ref{alg:new} providing
a more `targetted' way to use the simulated auxiliary variables. This
will be illustrated numerically in Section \ref{sec:examples}.

\subsubsection{Relating the variance of the estimator or $p_{\gamma}^{\epsilon}(y_{1:n})$
with the efficiency of ABC-MCMC}

A comparison of our results with the interesting work in \cite{pitt}
seems relevant. There the authors deal with a more general context
and  show that we should choose $N$ as a particular asymptotic (in $N$) variance; the main point is that 
the (asymptotic) variance of the estimate of $p_{\gamma}^{\epsilon}(y_{1:n})$ should be the same for each $\gamma$. We conjecture that in our set-up one
should choose $N$ such that the actual variance of the estimate of  $p_{\gamma}^{\epsilon}(y_{1:n})$ is constant with respect to $\gamma$. In this scenario, on inspection of the proof of Proposition \ref{prop:rel_var}, for a given $\gamma$,
one should set $N$ to be the solution of 
$$
\Big(\prod_{k=1}^n\alpha_k(y_{1:k},\epsilon,\gamma)^2\Big)\Big(\frac{1}{(N-1)^n(N-2)^n} - \frac{1}{(N-1)^{2n}}\Big) =C
$$
for some desired (upper-bound on the) variance $C$ (whose optimal value would need to be obtained).
This makes $N$ a random variable, in addition, but does not change the simulation mechanism. Unfortunately, one cannot do this in practice, as the $\alpha_k(y_{1:k},\epsilon,\gamma)$ are unknown. Taking into account Remark \ref{rem:gamma_indep}, this latter approach may not be so much of a concern in practice.

%

\subsubsection{On the ergodicity of the sampler}

We conclude this discussion by adding a related comment regarding
the ergodicity of the proposed MCMC kernel. If there exists a constant
$C<\infty$ such that 
\[
\frac{1}{\prod_{k=1}^{n}\alpha_{k}(y_{1:k},\epsilon,\gamma)}\leq C\quad\widehat{\pi}^{\epsilon}(d\gamma|y_{1:n})-a.e.
\]
and the marginal MCMC kernel in Algorithm \ref{alg:mcmc} is geometrically
ergodic, then by \cite[Propositions 7, 9]{andrieu1} the MCMC kernel
of Algorithm \ref{alg:new} is also geometrically ergodic.

\section{Examples}

\label{sec:examples}

\subsection{Scalar normal means model}

\subsubsection{Model}

For this example let each $Y_{k},X_{k},\theta$ be a scalar real random
variable and consider the model: 
\[
Y_{k+1}=\theta X_{k}+\kappa_{k},\quad X_{k+1}=X_{k}
\]
with $X_{0}=1$ and $\kappa_{k}\stackrel{\textrm{i.i.d.}}{\sim}\mathcal{N}(0,\sigma^{2})$,
where we denote $\mathcal{N}(0,\sigma^{2})$ the zero mean normal
distribution with variance $\sigma^{2}$. The prior on $\theta$ is
$\mathcal{N}(0,\phi)$. This model is usually referred to as the standard
normal means model in one dimension and the posterior is given by:
\[
\theta|y_{1:n}\sim\mathcal{N}\Big(\frac{\sigma_{n}^{2}}{\sigma^{2}}\sum_{k=1}^{n}y_{k},\sigma_{n}^{2}\Big)
\]
where $\sigma_{n}^{2}=\Big(\frac{1}{\phi}+\frac{n}{\sigma^{2}}\Big)^{-1}.$
Note that if $Y_{k}\stackrel{\textrm{i.i.d.}}{\sim}\mathcal{N}(\theta^{*},\sigma^{2})$,
then the posterior on $\theta$ is consistent and concentrates around
$\theta^{*}$  as $n\rightarrow\infty$.

The ABC approximation after marginalizing out the auxiliary variables
has a likelihood given by: 
\[
p_{\theta}^{\epsilon}(y_{1:n})=\frac{1}{\epsilon^{n}}\prod_{k=1}^{n}\Big[F\Big(\frac{y_{k}+\epsilon-\theta}{\sigma}\Big)-F\Big(\frac{y_{k}-\epsilon-\theta}{\sigma}\Big)\Big]
\]
 where $F$ is the standard normal cumulative density function.
Thus, this is a scenario where we can perform the marginal MCMC.

\subsubsection{Simulation Results}

Three data sets are generated from the model with $n\in\{10,100,1000\}$
and $\sigma^{2}=1$. In addition, for $\epsilon=1$ we perturb the
data-sets in order to use them for noisy ABC. For the sake of comparison,
we also generate a noisy ABC data-set for $\epsilon=100$. We will
also use a prior with $\phi=1$. 

\begin{figure}[h]
\centering \subfigure[$n=10,\epsilon=1$]{{\includegraphics[width=0.49\textwidth,height=7cm]{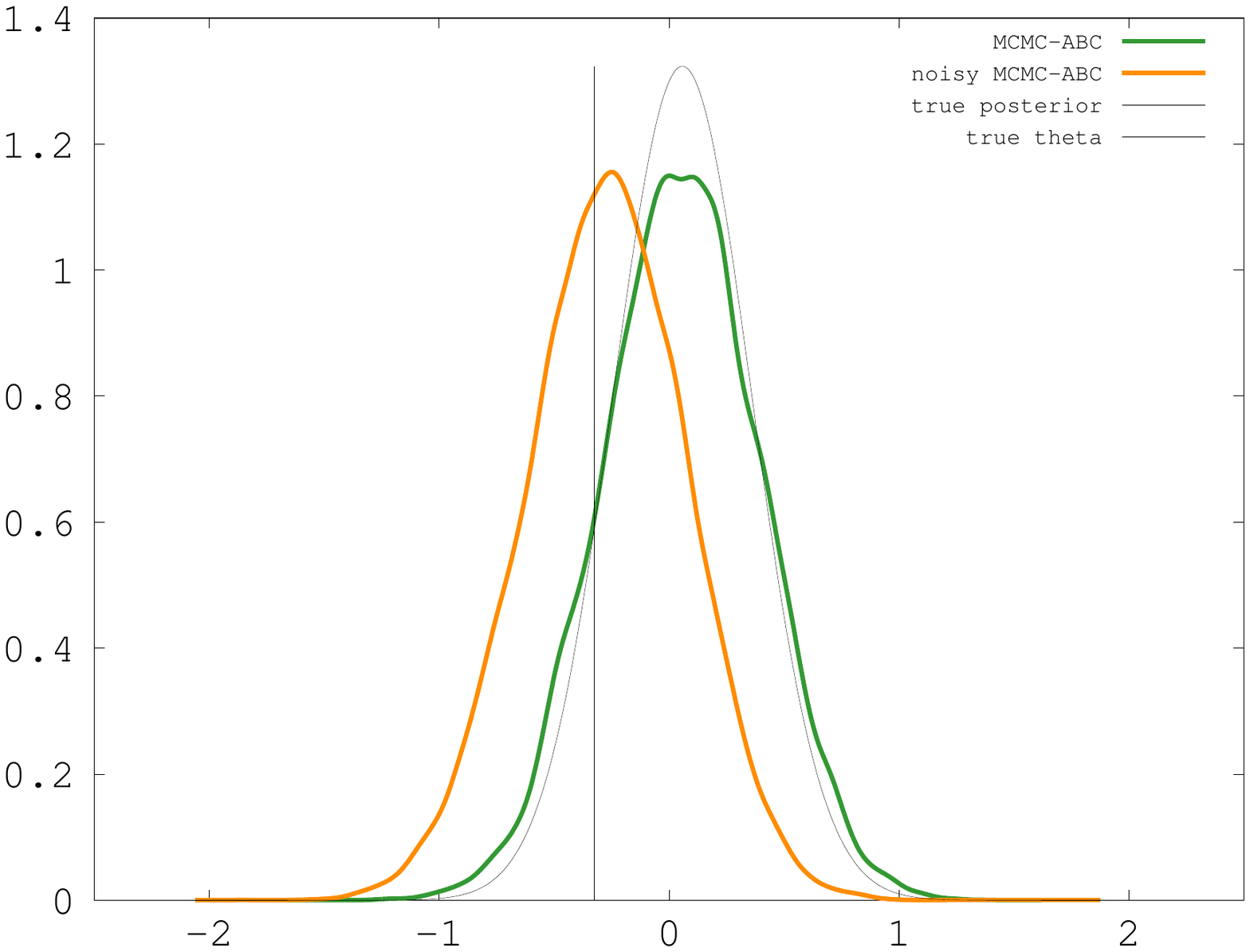}}}
\subfigure[$n=10,\epsilon=100$]{{\includegraphics[width=0.49\textwidth,height=7cm]{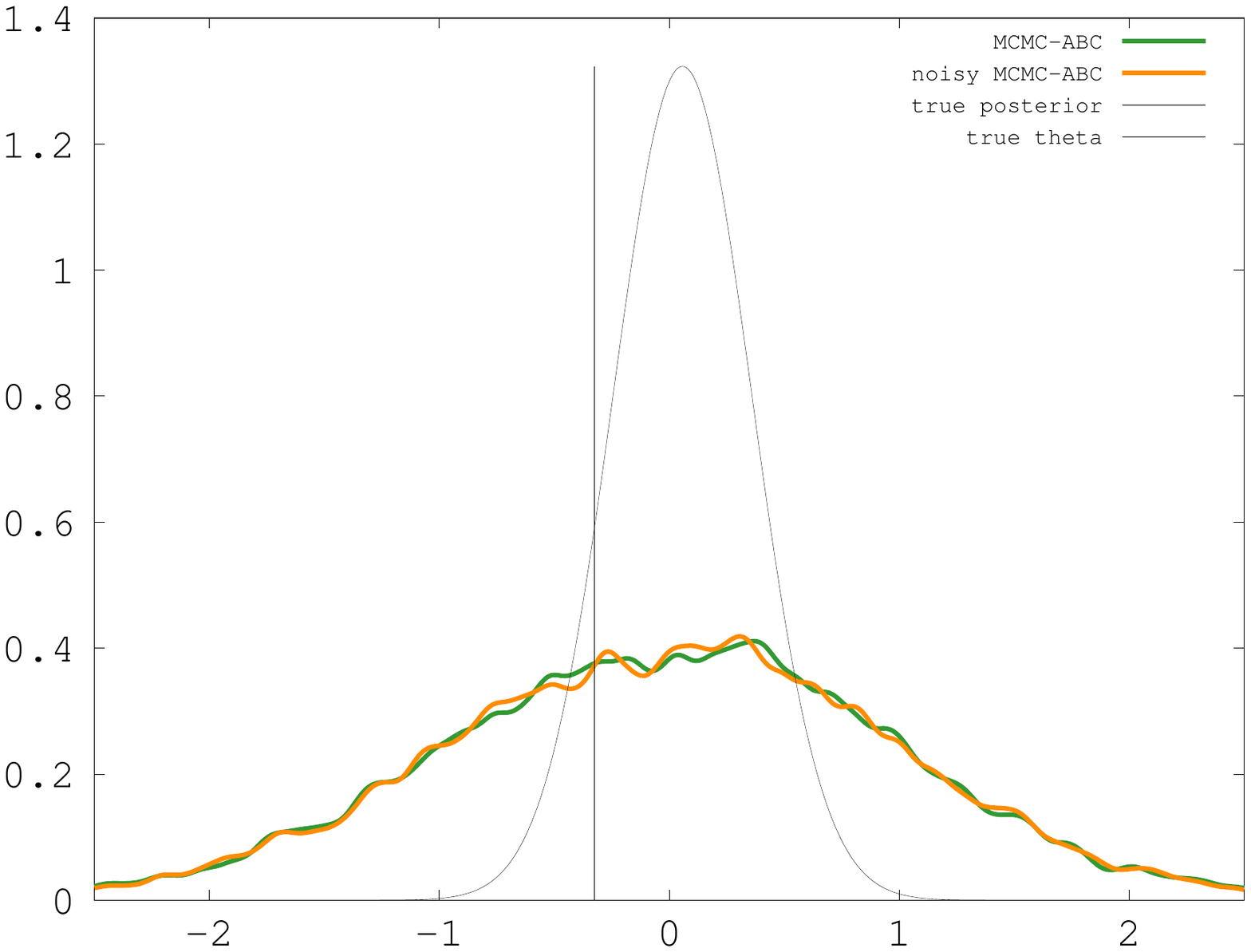}}}
\subfigure[$n=100,\epsilon=1$]{{\includegraphics[width=0.49\textwidth,height=7cm]{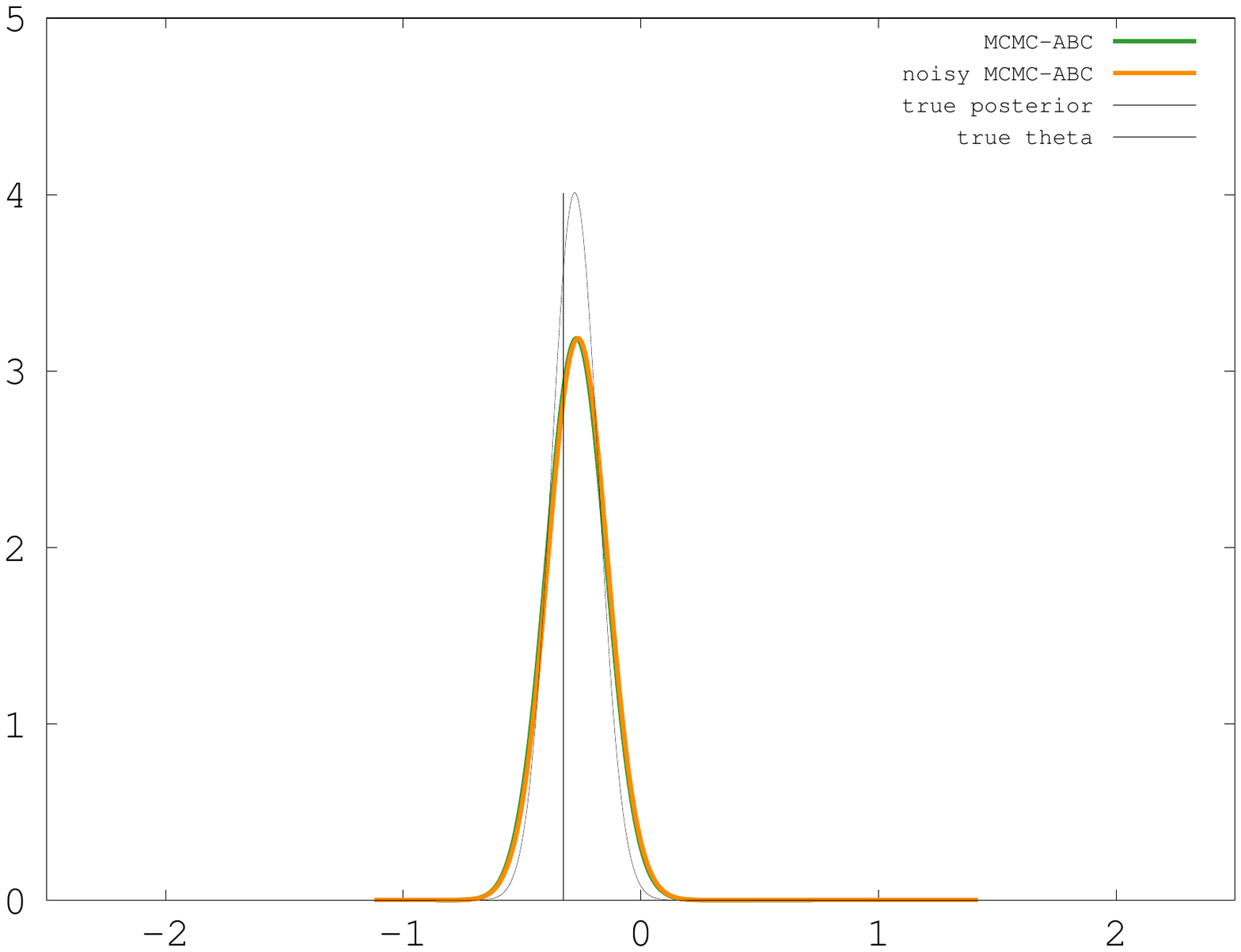}}}
\subfigure[$n=100,\epsilon=100$]{{\includegraphics[width=0.49\textwidth,height=7cm]{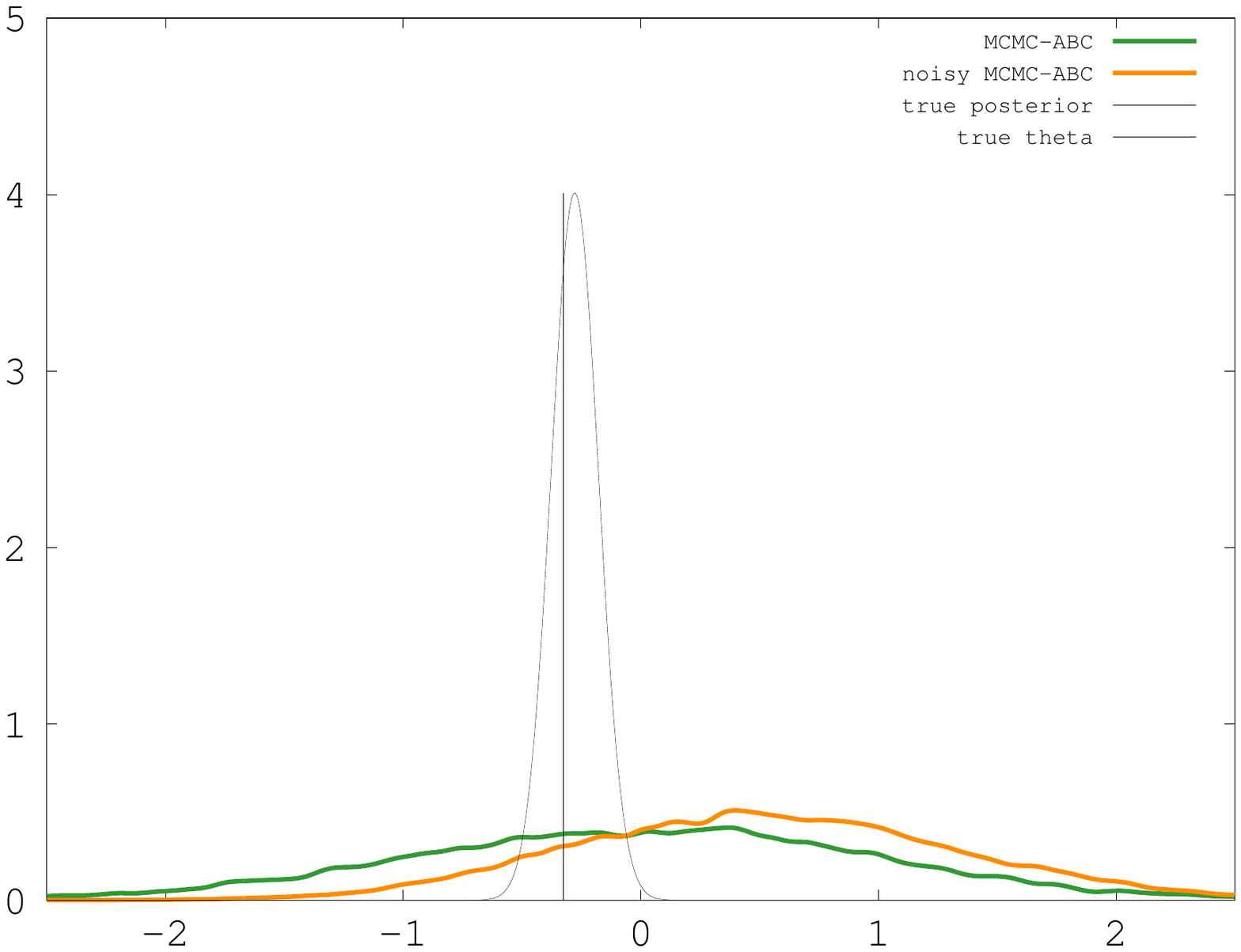}}}
\subfigure[$n=1000,\epsilon=1$]{{\includegraphics[width=0.49\textwidth,height=7cm]{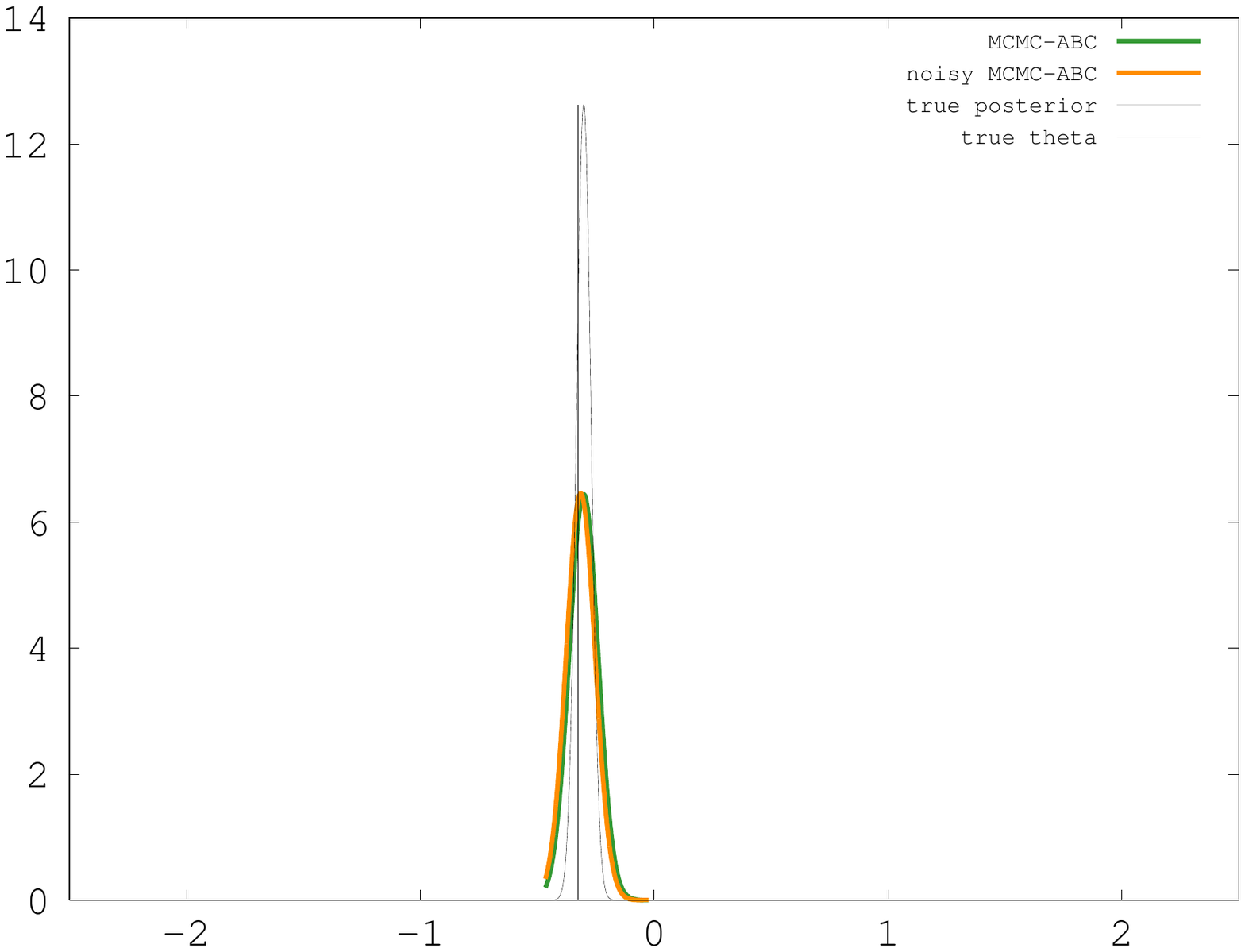}}}
\subfigure[$n=1000,\epsilon=100$]{{\includegraphics[width=0.49\textwidth,height=7cm]{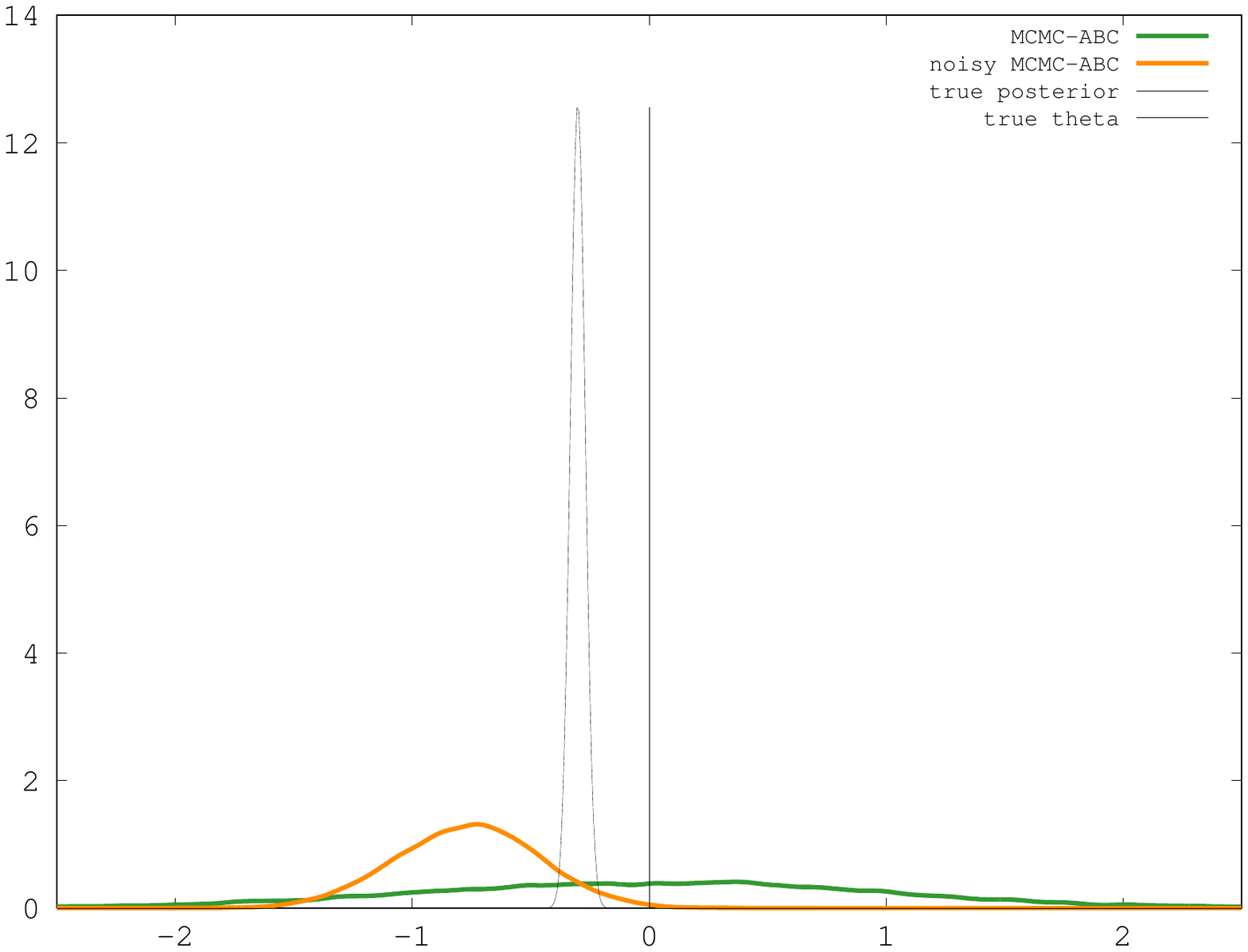}}}
\caption{Marginal MCMC Density Plots. In each plot the true posterior (black),
noisy ABC (orange), ABC (green) densities of $\theta$ are plotted,
for different values of $n$ (10, 1st row, 100, 2nd row, 1000, 3rd
row) and $\epsilon$ (1, 1st column, 100, 2nd column). The black vertical
line is the value of $\theta$ that generated the data.}

\label{fig:marginal_den_normal} 
\end{figure}

\begin{figure}[h]
\centering \subfigure[$n=10$, ABC]{{\includegraphics[width=0.49\textwidth,height=7cm]{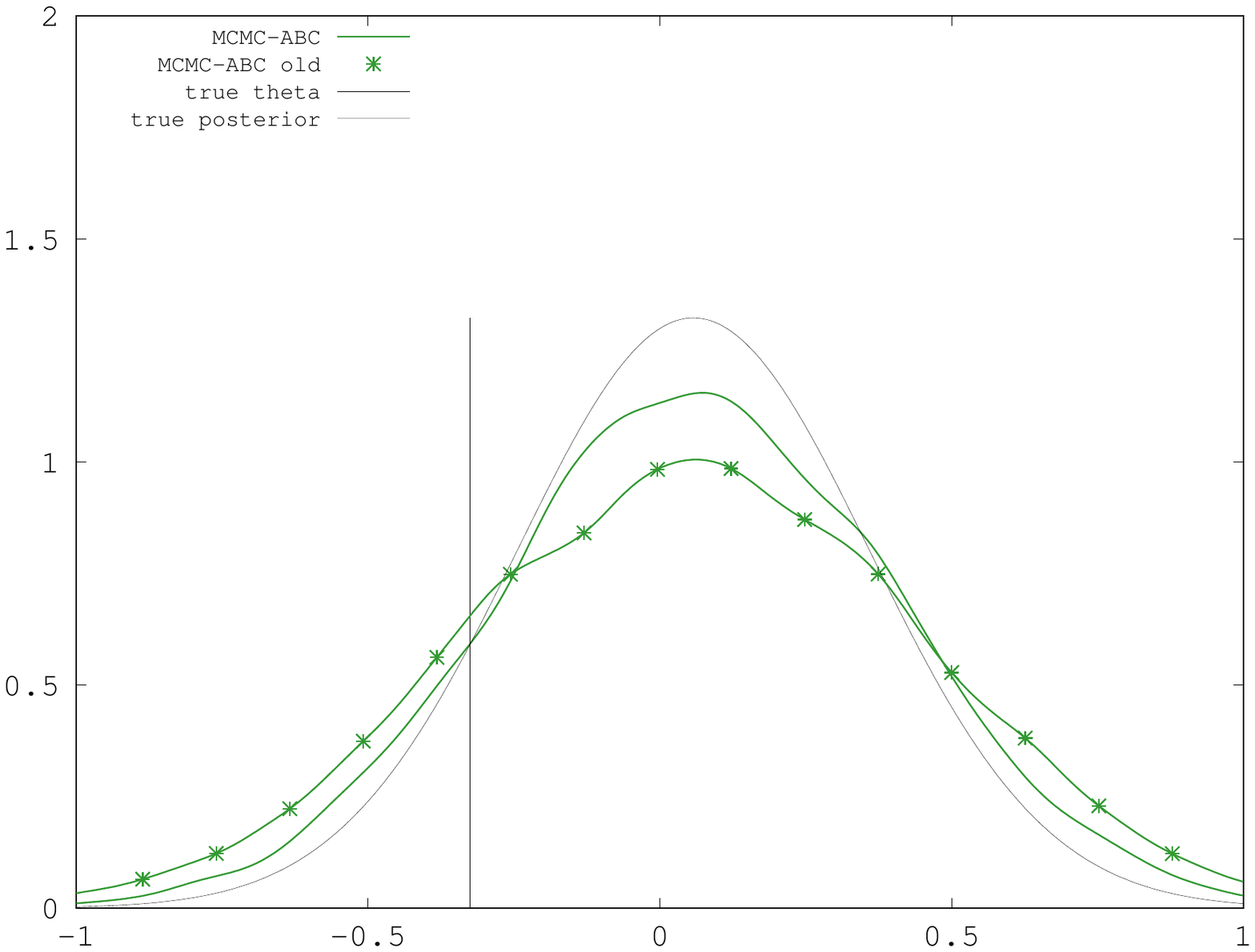}}}
\subfigure[$n=10$, Noisy ABC]{{\includegraphics[width=0.49\textwidth,height=7cm]{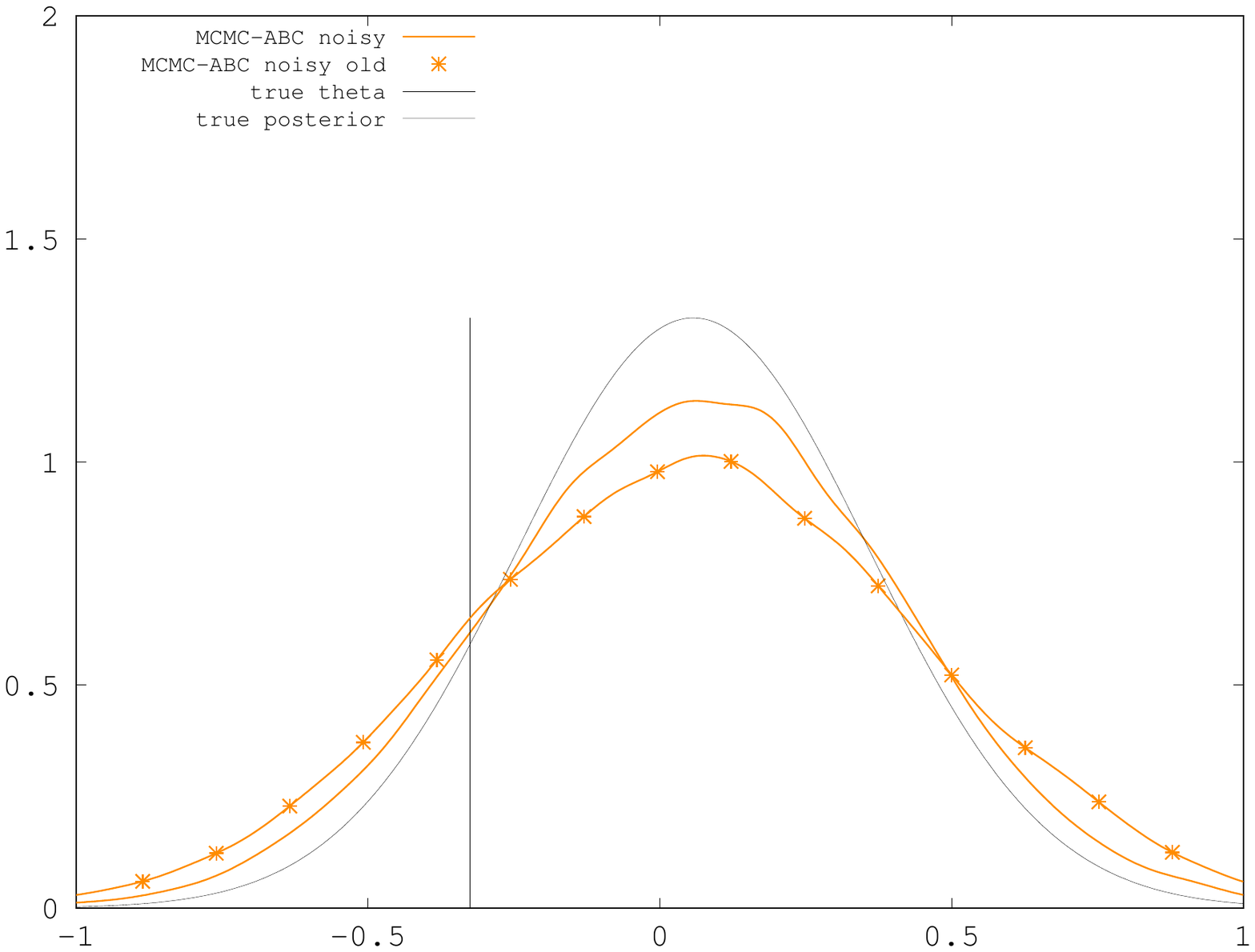}}}
\subfigure[$n=100$, ABC]{{\includegraphics[width=0.49\textwidth,height=7cm]{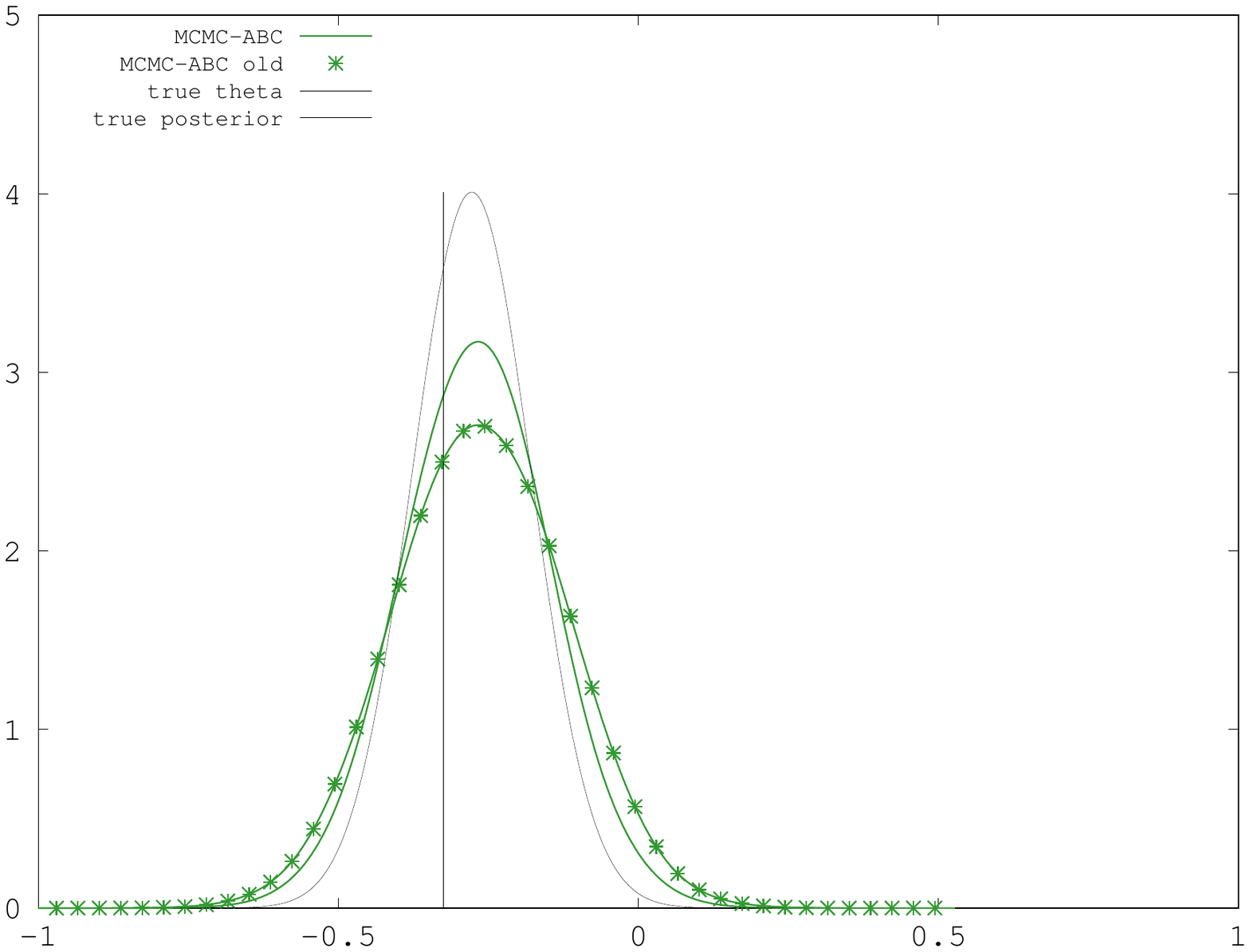}}}
\subfigure[$n=100$, Noisy ABC]{{\includegraphics[width=0.49\textwidth,height=7cm]{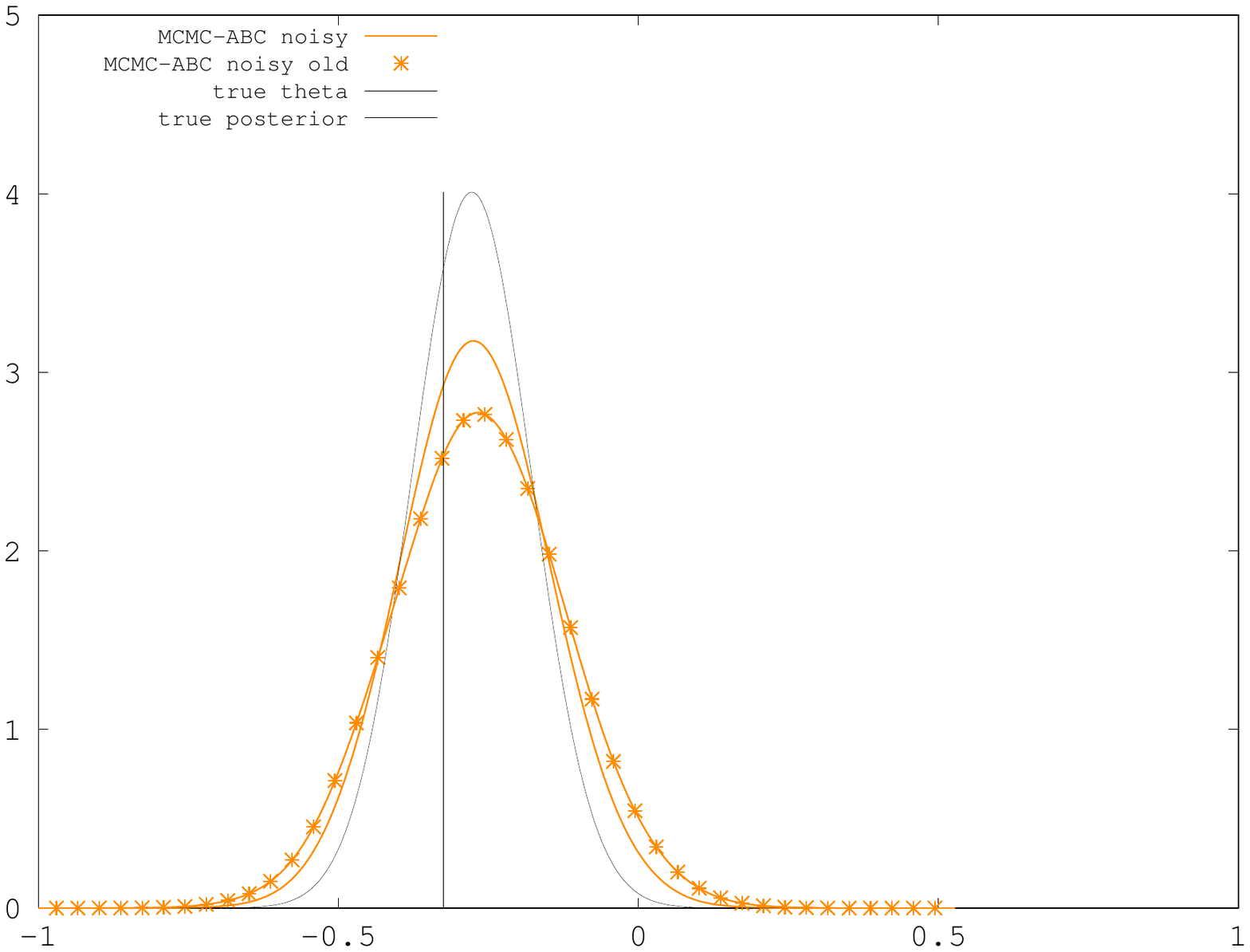}}}
\subfigure[$n=1000$, ABC]{{\includegraphics[width=0.49\textwidth,height=7cm]{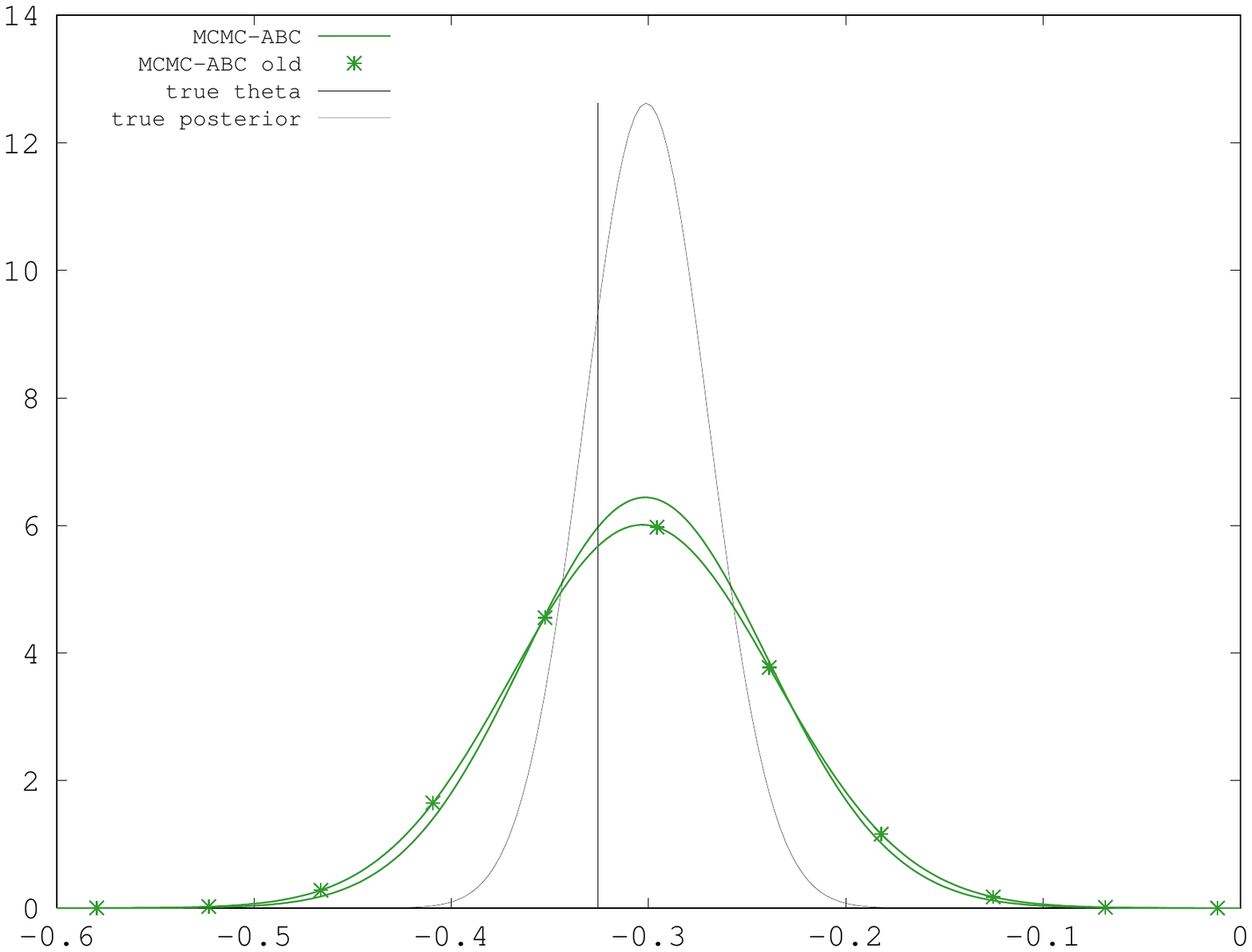}}}
\subfigure[$n=1000$, Noisy ABC]{{\includegraphics[width=0.49\textwidth,height=7cm]{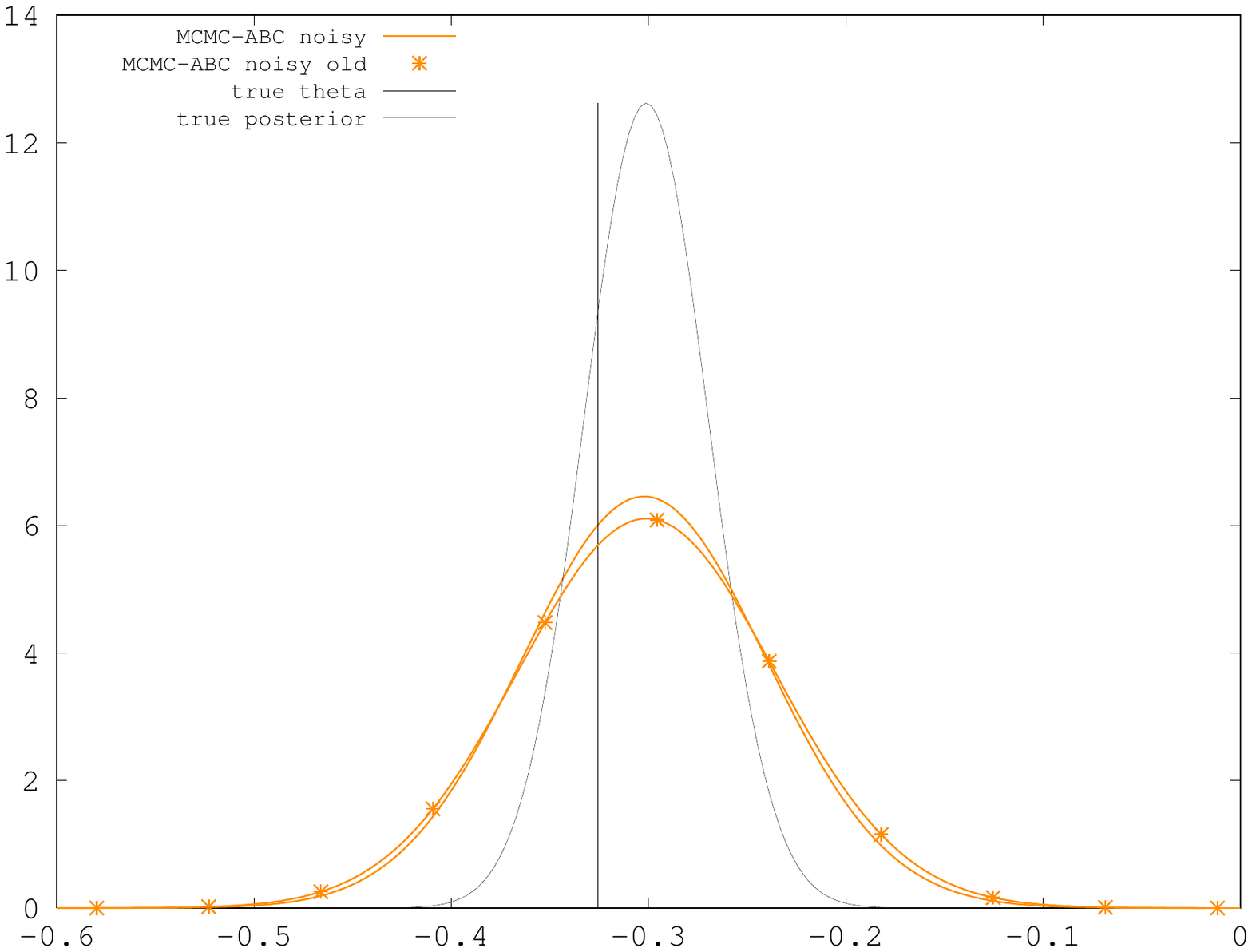}}}
\caption{MCMC Density Plots. In each plot the true posterior (black), ABC (green,
1st col) or noisy ABC (orange, 2nd col) densities of $\theta$ are
plotted, for different values of $n$ (10, 1st row, 100, 2nd row,
1000, 3rd row). In addition the plots are for the new and old (the
{*}) MCMC kernels. The black vertical line is the value of $\theta$
that generated the data. Throughout $\epsilon=1$}

\label{fig:marginal_den_oldnew_normal} 
\end{figure}

\begin{figure}[h]
\centering \subfigure[$n=10$, Marginal]{{\includegraphics[width=0.49\textwidth,height=6.5cm]{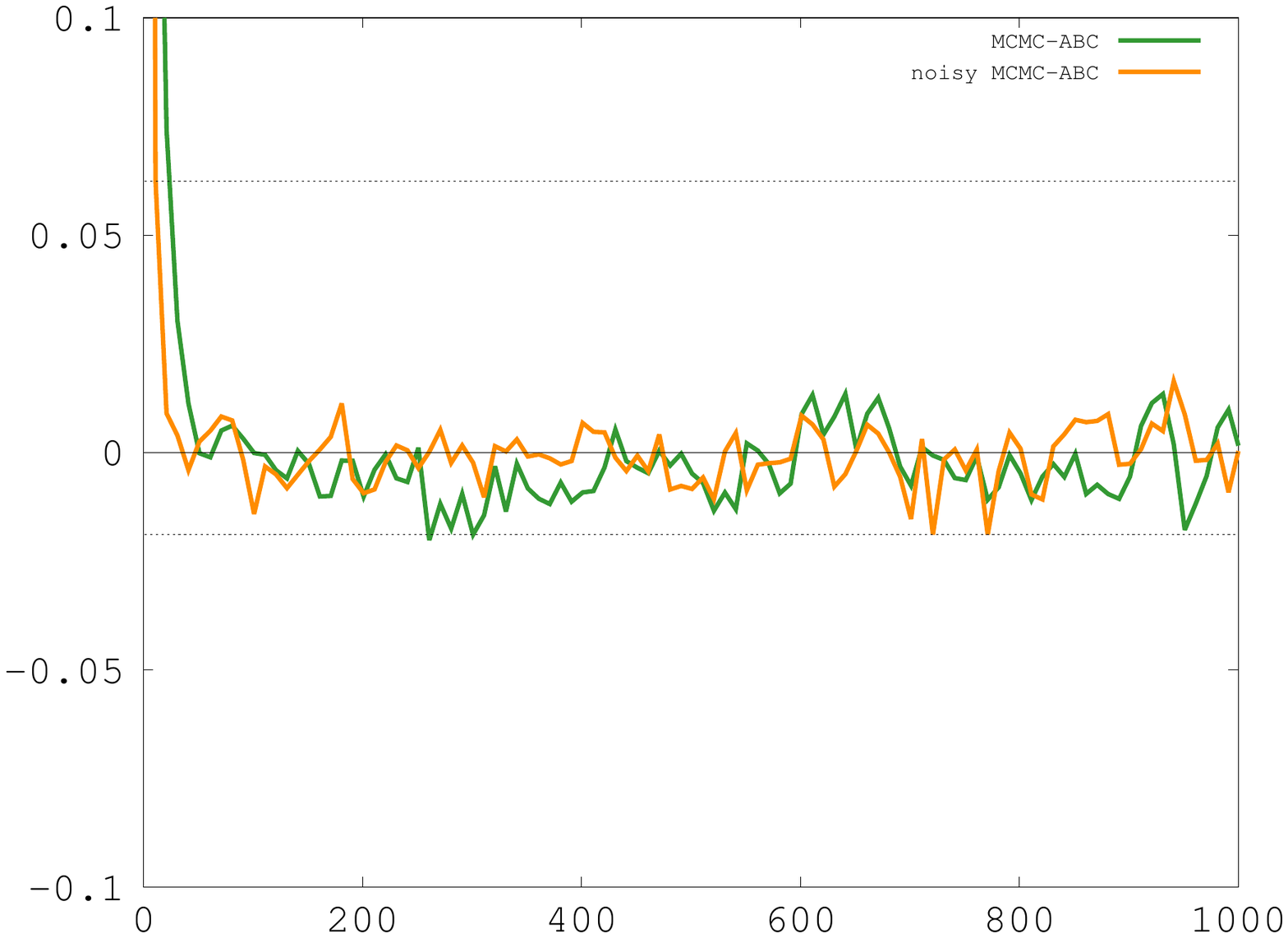}}}
\subfigure[$n=10$, Standard]{{\includegraphics[width=0.49\textwidth,height=6.5cm]{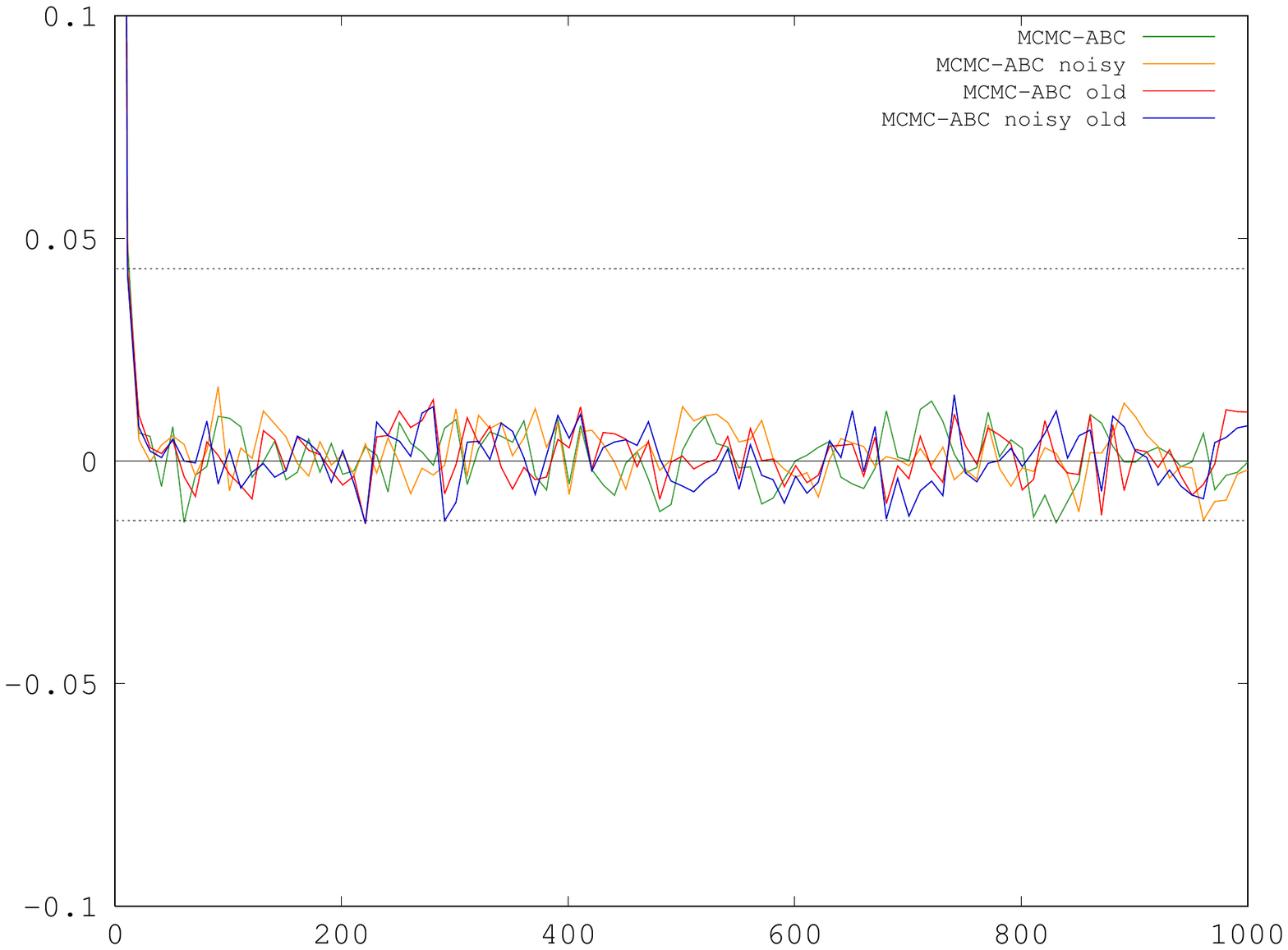}}}
\subfigure[$n=100$, Marginal]{{\includegraphics[width=0.49\textwidth,height=6.5cm]{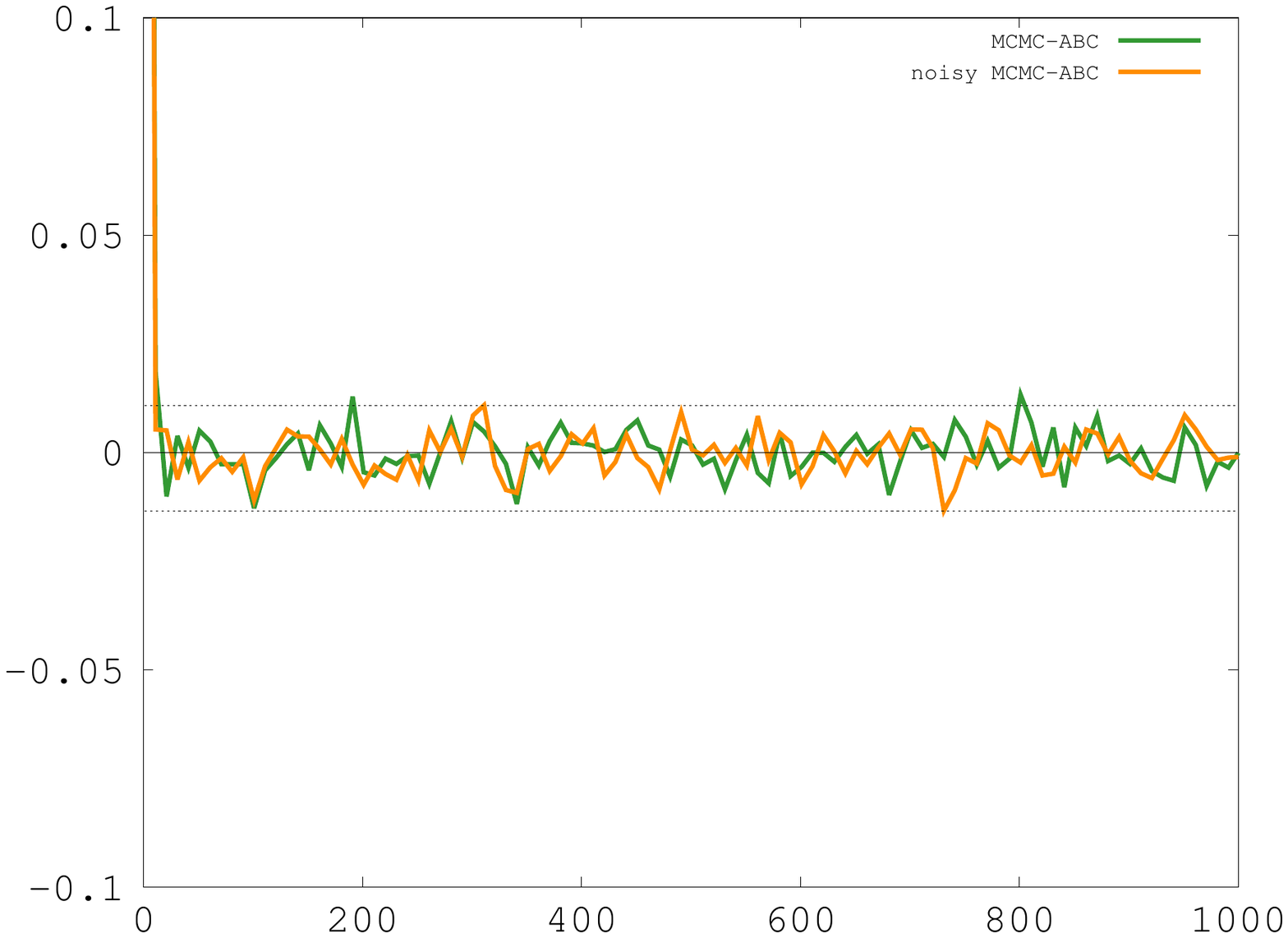}}}
\subfigure[$n=100$, Standard]{{\includegraphics[width=0.49\textwidth,height=6.5cm]{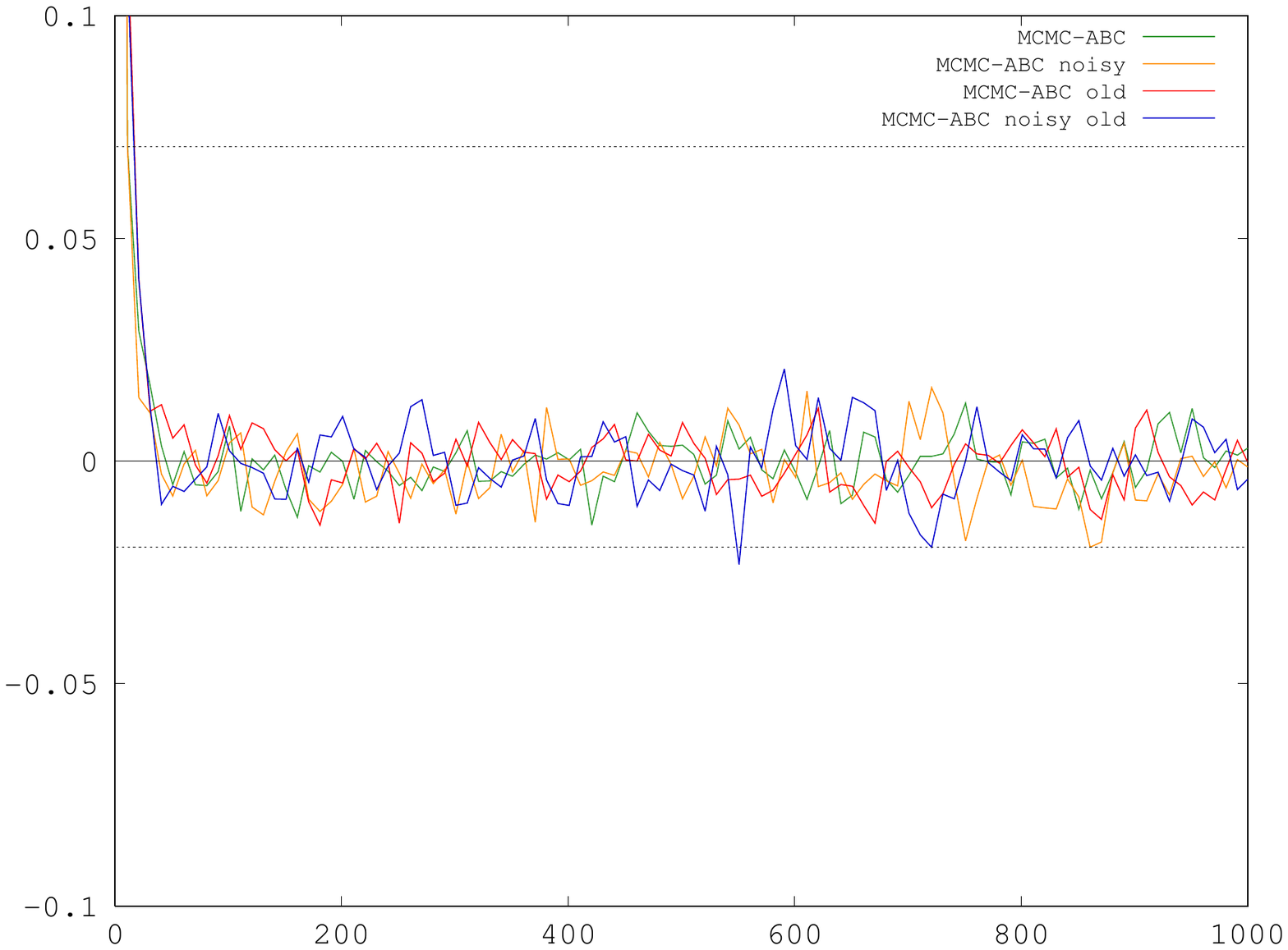}}}
\subfigure[$n=1000$, Marginal]{{\includegraphics[width=0.49\textwidth,height=6.5cm]{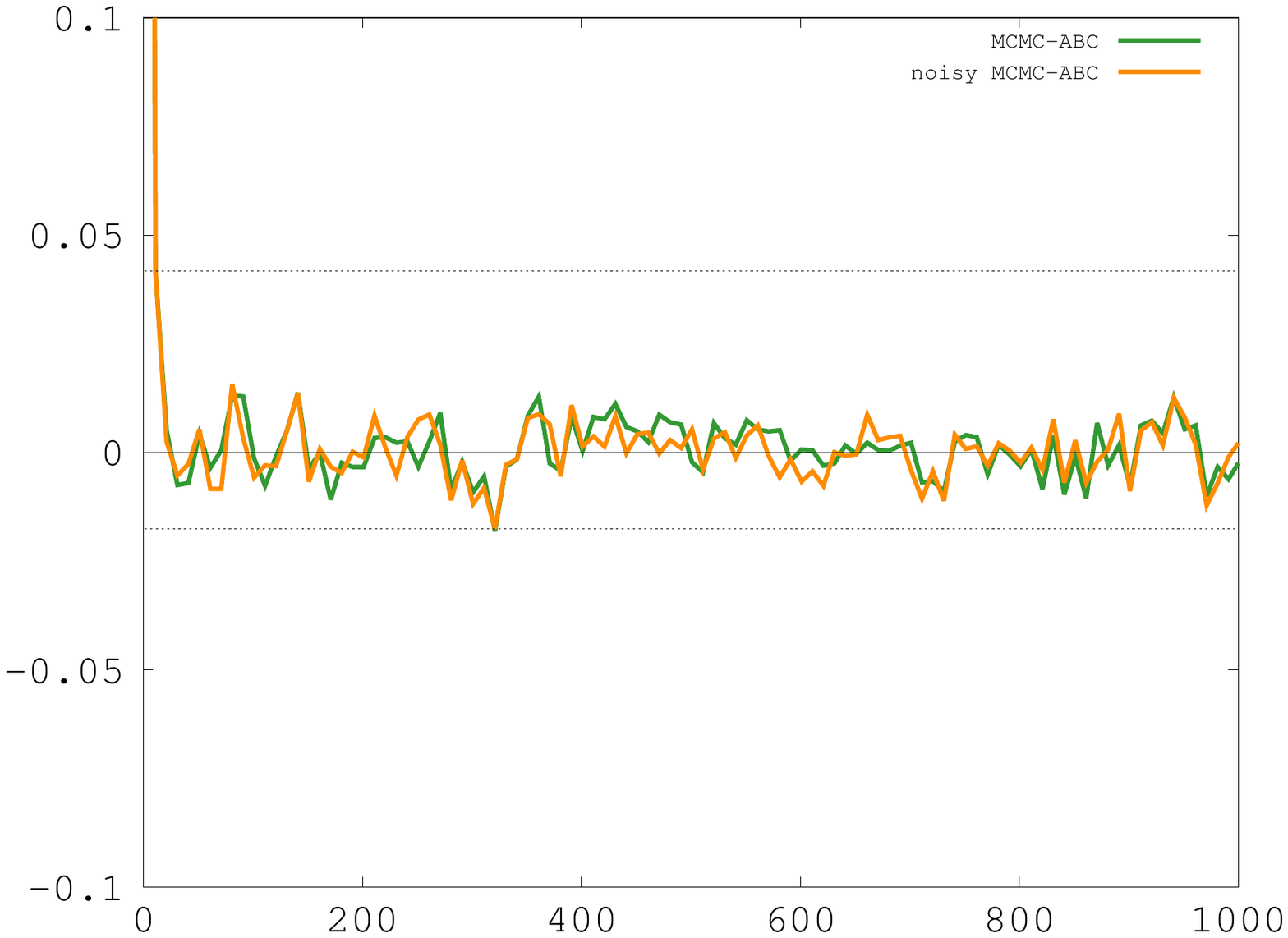}}}
\subfigure[$n=1000$, Standard]{{\includegraphics[width=0.49\textwidth,height=6.5cm]{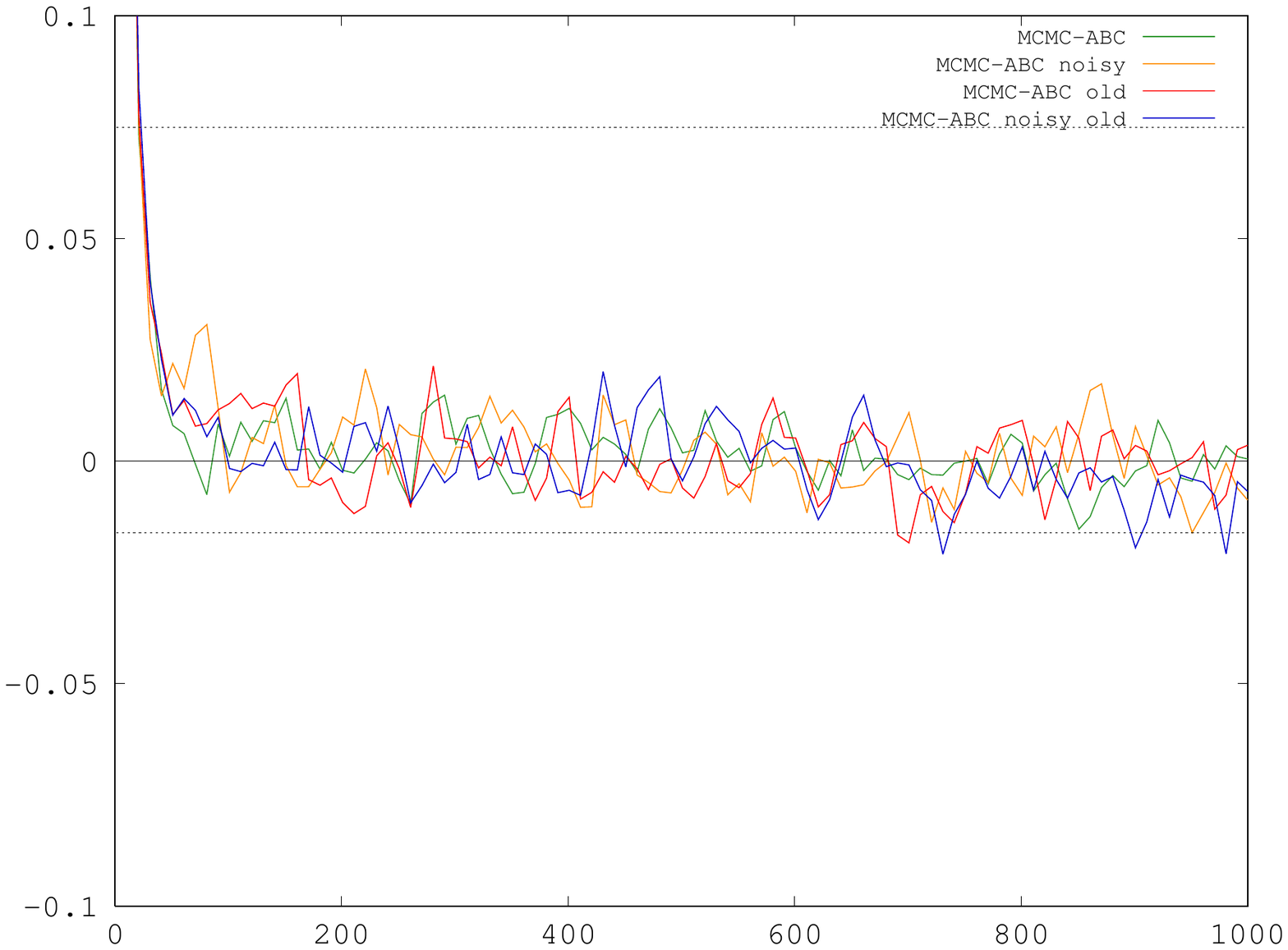}}}
\caption{Marginal MCMC and Standard Auto-Correlation Plots. In each plot in
column 1 the auto-correlation for every 10th iteration is plotted
for noisy ABC (orange) and ABC (green). In each plot in column 2 the
auto-correlation over the same period is plotted for both the new
and old MCMC kernels, for both noisy ABC (orange for new and blue
for old) and ABC (green for new, red for old). Three different values
of $n$ are presented and $\epsilon=1$ throughout.}

\label{fig:marginal_acf_normal} 
\end{figure}

We run the new MCMC kernel (the proposal in Algorithm \ref{alg:new} - we will frequently use the expression `Algorithm' to mean an MCMC kernel with the given proposal mechansim of the
Algorithm), old
MCMC kernel (Algorithm \ref{alg:Ntry})
and a Marginal MCMC algorithm which just samples on the parameter
space $\mathbb{R}$ (i.e.~the posterior density is proportional to
$p_{\theta}^{\epsilon}(y_{1:n})\pi(\theta)$). Each algorithm is run
with a normal random walk proposal on the parameter space, with the
same scaling. The scaling chosen yields an acceptance rate of around
0.25 for each run of the marginal MCMC algorithm. The new MCMC kernel
is run with $N=n$ and the old with a slightly higher value of $N$
so that the computational times are about the same (so for example,
the running time of the new kernel is not a problem in this example).
The algorithms are run for 10000 iterations and the results can be
found in Figures \ref{fig:marginal_den_normal}-\ref{fig:marginal_acf_normal}.

In Figure \ref{fig:marginal_den_normal} the density plots for the
posterior samples on $\theta$, from the marginal MCMC can be seen
for $\epsilon\in\{1,100\}$ and each value of $n$. When $\epsilon=1$,
we can observe that both ABC and noisy ABC both get closer to the
true posterior as $n$ grows. For noisy ABC, this is the behavior
that is predicted in Section \ref{sec:consis}. For the ABC approximation,
following the proof of Theorem 1 in \cite{jasra}, one can see that
the bias falls with $\epsilon$; hence, in this scenario there is
not a substantial bias for the standard ABC approximation. When we
make $\epsilon$ much larger a more pronounced difference between
ABC and noisy ABC can be seen and it appears as $n$ grows that the
noisy ABC approximation is slightly more accurate (relative to ABC).

We now consider the similarity of the new and old MCMC kernels to
the marginal algorithm (i.e.~the kernel both procedures attempt to
approximate), the results are in Figures \ref{fig:marginal_den_oldnew_normal}-\ref{fig:marginal_acf_normal}.
With regards to both the density plots (Figure \ref{fig:marginal_den_oldnew_normal})
and auto-correlations (Figure \ref{fig:marginal_acf_normal}) we can
see that both MCMC kernels appear to be quite similar to the marginal
MCMC. It is also noted that the acceptance rates of these latter kernels
are also not far from that of the marginal algorithm (results not
shown). These results are unsuprising, given the simplicity of the
density that we target, but still reassuring; a more comprehensive
comparison is given in the next example. Encouragingly, the new and
old MCMC kernels do not seem to noticably worsen as $n$ grows; this
shows that, at least for this example, the recommendation of $N=\mathcal{O}(n)$
is quite useful. We remark that whilst these results are for a single
batch of data, the results are consistent with other data sets.

\subsection{Real Data Example}

\subsubsection{Model}

Set, for $(Y_{k},X_{k})\in\mathbb{R}\times\mathbb{R}^{+}$ 
\begin{eqnarray*}
Y_{k+1} & = & \kappa_{k}\quad k\in\mathbb{N}_{0}\\
X_{k+1} & = & \beta_{0}+\beta_{1}X_{k}+\beta_{2}Y_{k+1}^{2}\quad k\in\mathbb{N}_{0}
\end{eqnarray*}
 where $\kappa_{k}|x_{k}\stackrel{\textrm{ind}}{\sim}\mathcal{S}(0,x_{k},\varphi_{1},\varphi_{2})$
(i.e.~a stable distribution, with location 0, scale $X_{k}$ and
asymmetry and skewness parameters $\varphi_{1},\varphi_{2}$). We
set 
\[
X_{0}\sim\mathcal{G}a(a,b),\quad\beta_{0},\beta_{1},\beta_{2}\sim\mathcal{G}a(c,d)
\]
 where $\mathcal{G}a(a,b)$ is a Gamma distribution with mean $a/b$
and $\theta=(\beta_{0:2})\in(\mathbb{R}^{+})^{3}$. This is a GARCH(1,1)
model with an intractable likelihood.

\subsubsection{Simulation Results}

We consider daily log-returns data from the S\&P 500 index from 03/1/11 to 14/02/13, which 
constitutes 533 data-points. In the priors, we set $a=c=2$ and $b=d=1/8$, which are not overly informative.
In addition, $\varphi_1=1.5$ and $\varphi_2=0$.
We consider $\epsilon\in\{0.01,0.5\}$ and only a noisy ABC approximation of the model.
Algorithms \ref{alg:Ntry} and \ref{alg:new} are to be compared.
 The MCMC proposals on the parameters are random-walks on the log-scale and for both algorithms we set $N=250$. 
It should be noted that our results are fairly robust to changes in $N\in[100,500]$, which are the values we tested the algorithm with.

\begin{figure}[h]
\centering \subfigure[$X_0$]{{\includegraphics[width=0.49\textwidth,height=6.5cm]{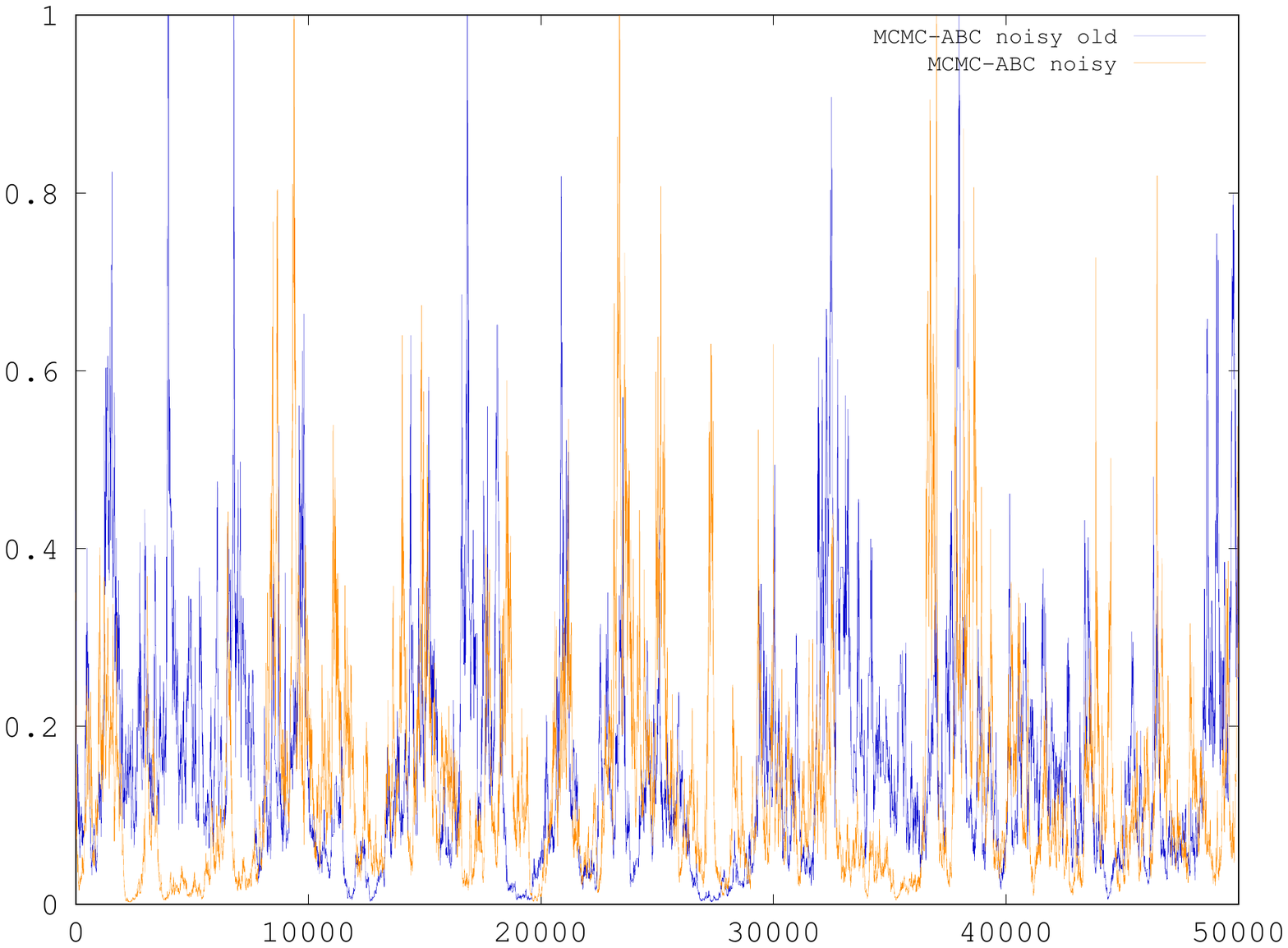}}}
\subfigure[$\beta_0$]{{\includegraphics[width=0.49\textwidth,height=6.5cm]{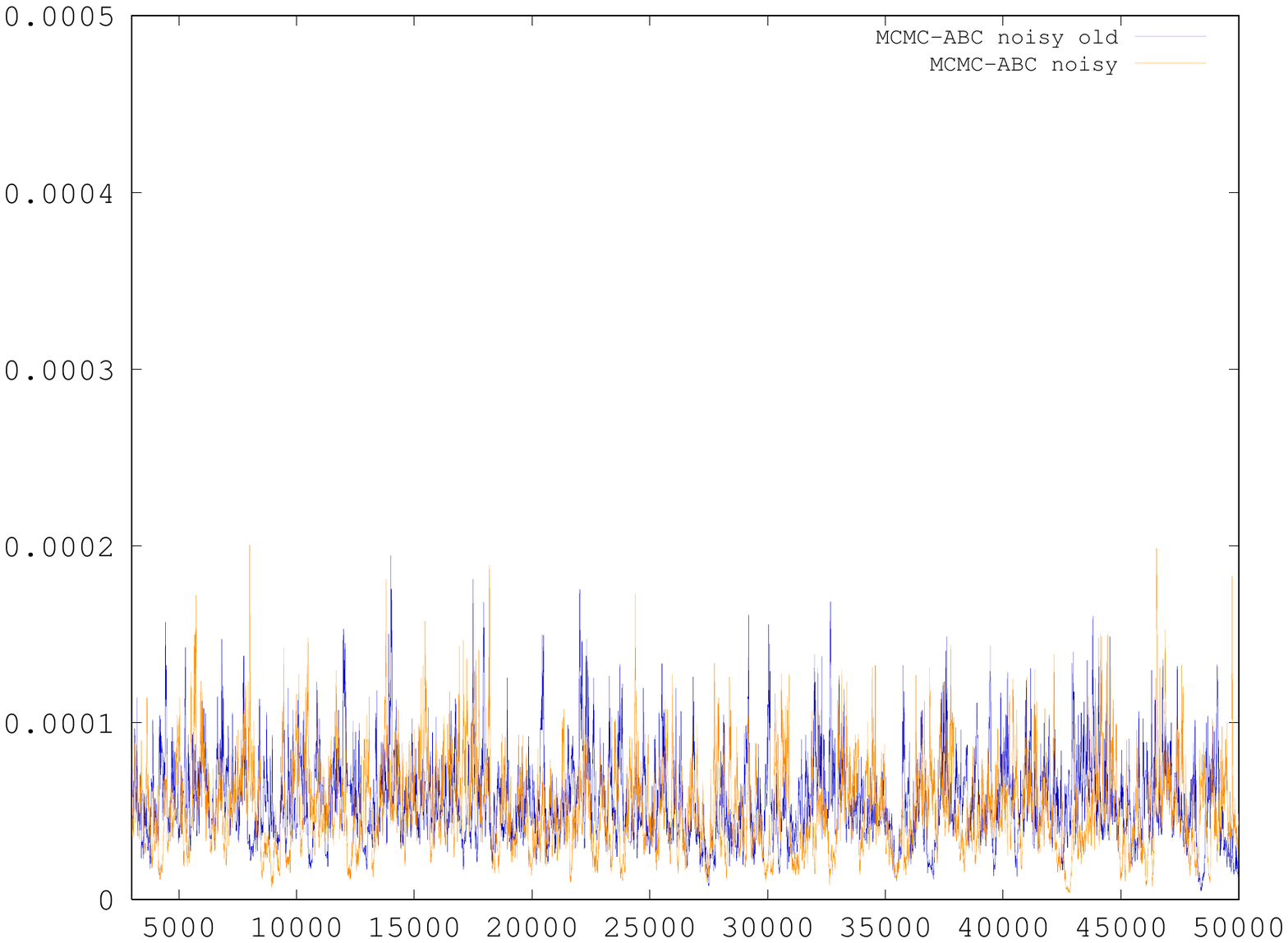}}}
\subfigure[$\beta_1$]{{\includegraphics[width=0.49\textwidth,height=6.5cm]{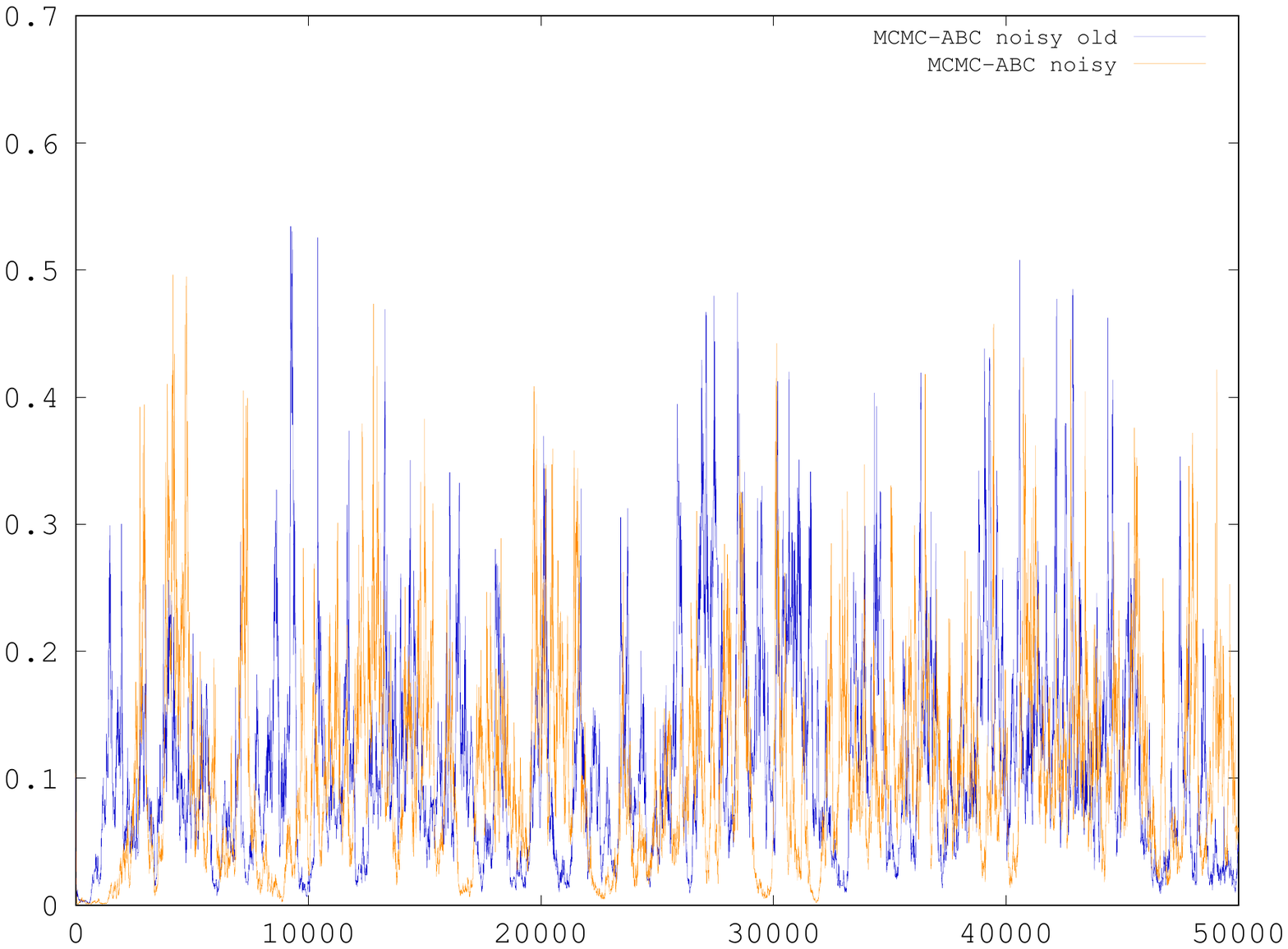}}}
\subfigure[$\beta_2$]{{\includegraphics[width=0.49\textwidth,height=6.5cm]{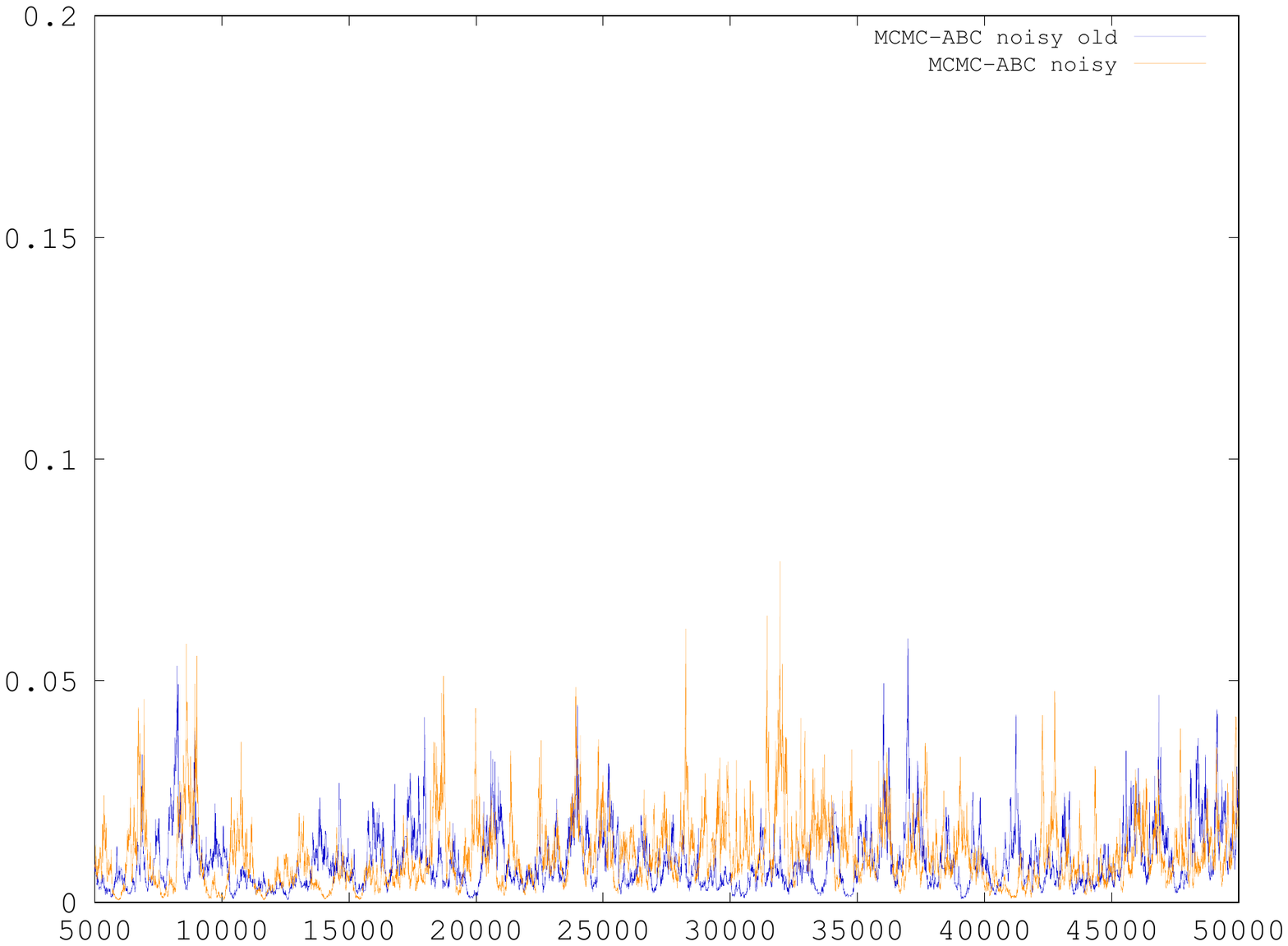}}}
\caption{Trace, $\epsilon=0.5$. We run both algorithms for 50000 iterations, the new kernel is in orange trace and $N=250$ in both cases and the algorithms
are run the S \& P 500 data.}
\label{fig:tracee05} 
\end{figure}

In Figure \ref{fig:tracee05} we present the trace-plot of 50000 iterations of both MCMC kernels when $\epsilon=0.5$. 
Algorithm \ref{alg:Ntry} took about 0.30 seconds per iteration and Algorithm \ref{alg:new} took about 1.12 seconds per iteration
We modified the proposal variances to yield an acceptance rate around 0.3.
The plot shows that both algorithms appear to move across the state-space in a very reasonable way. The new algorithm takes much longer and in this situation does not appear to be required. This run is one of many we performed and we observed this behaviour in many of our runs.

\begin{figure}[h]
\centering \subfigure[$X_0$]{{\includegraphics[width=0.49\textwidth,height=3cm]{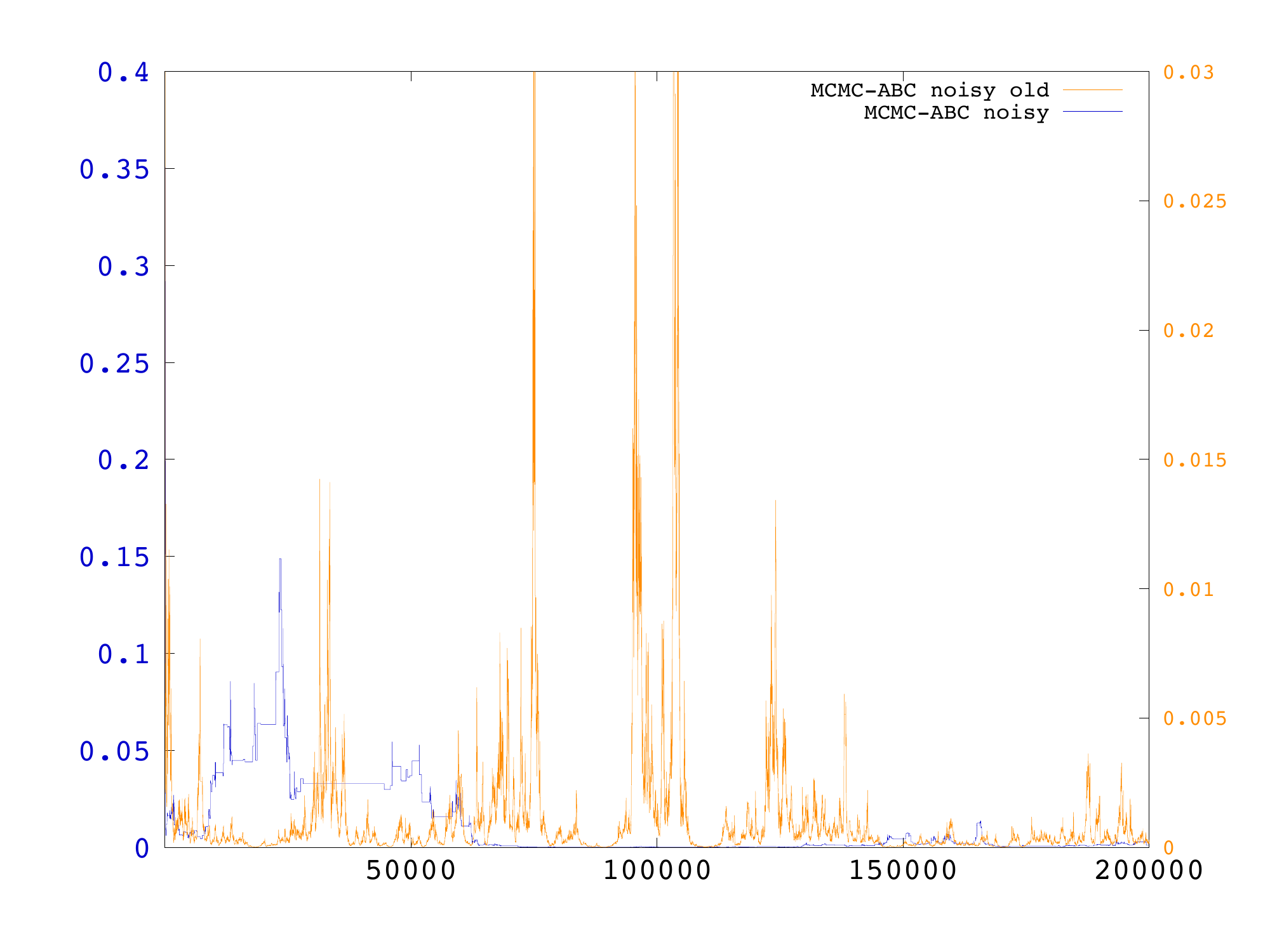}}}
\subfigure[$\beta_0$]{{\includegraphics[width=0.49\textwidth,height=3cm]{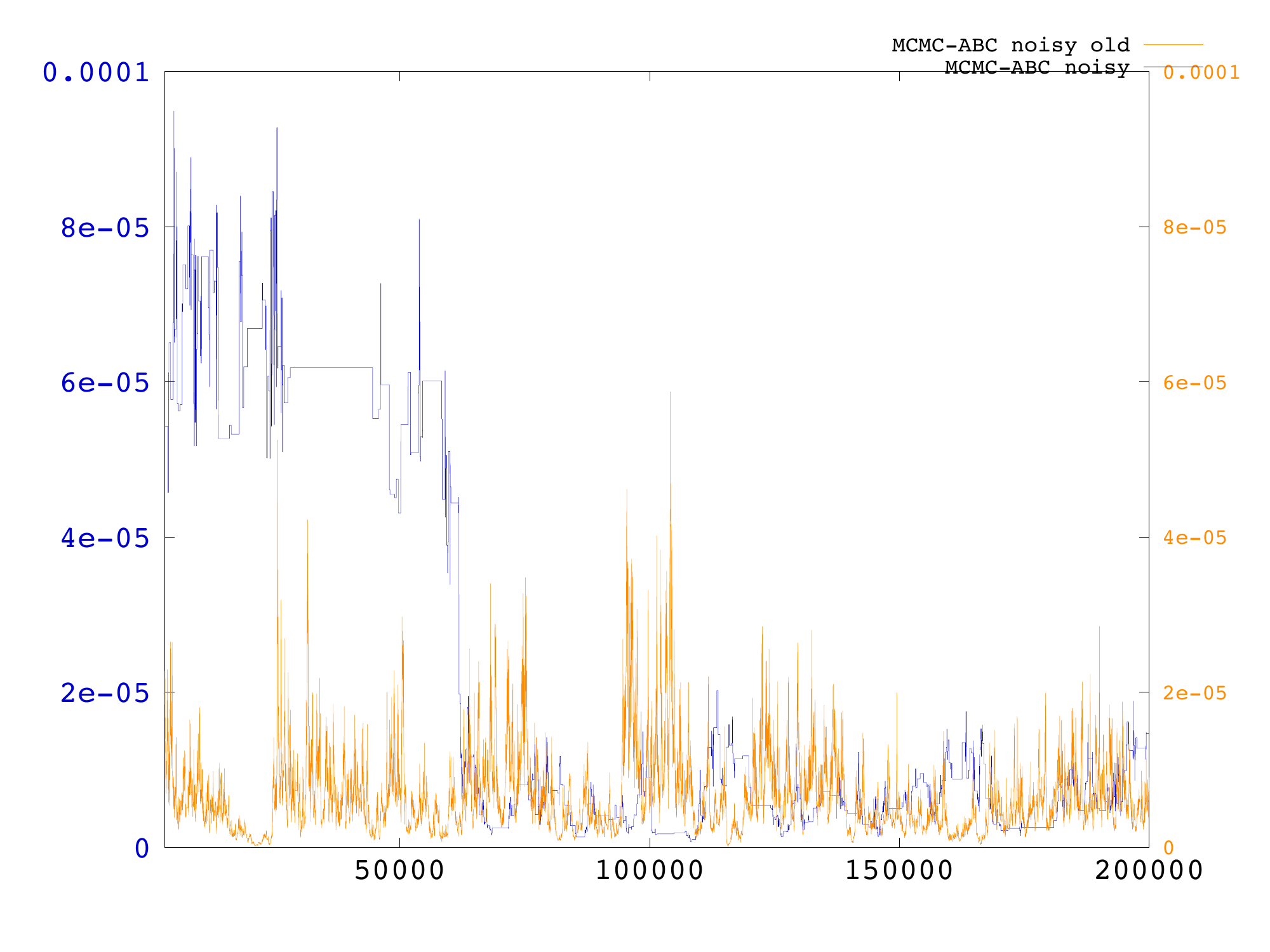}}}
\subfigure[$\beta_1$]{{\includegraphics[width=0.49\textwidth,height=3cm]{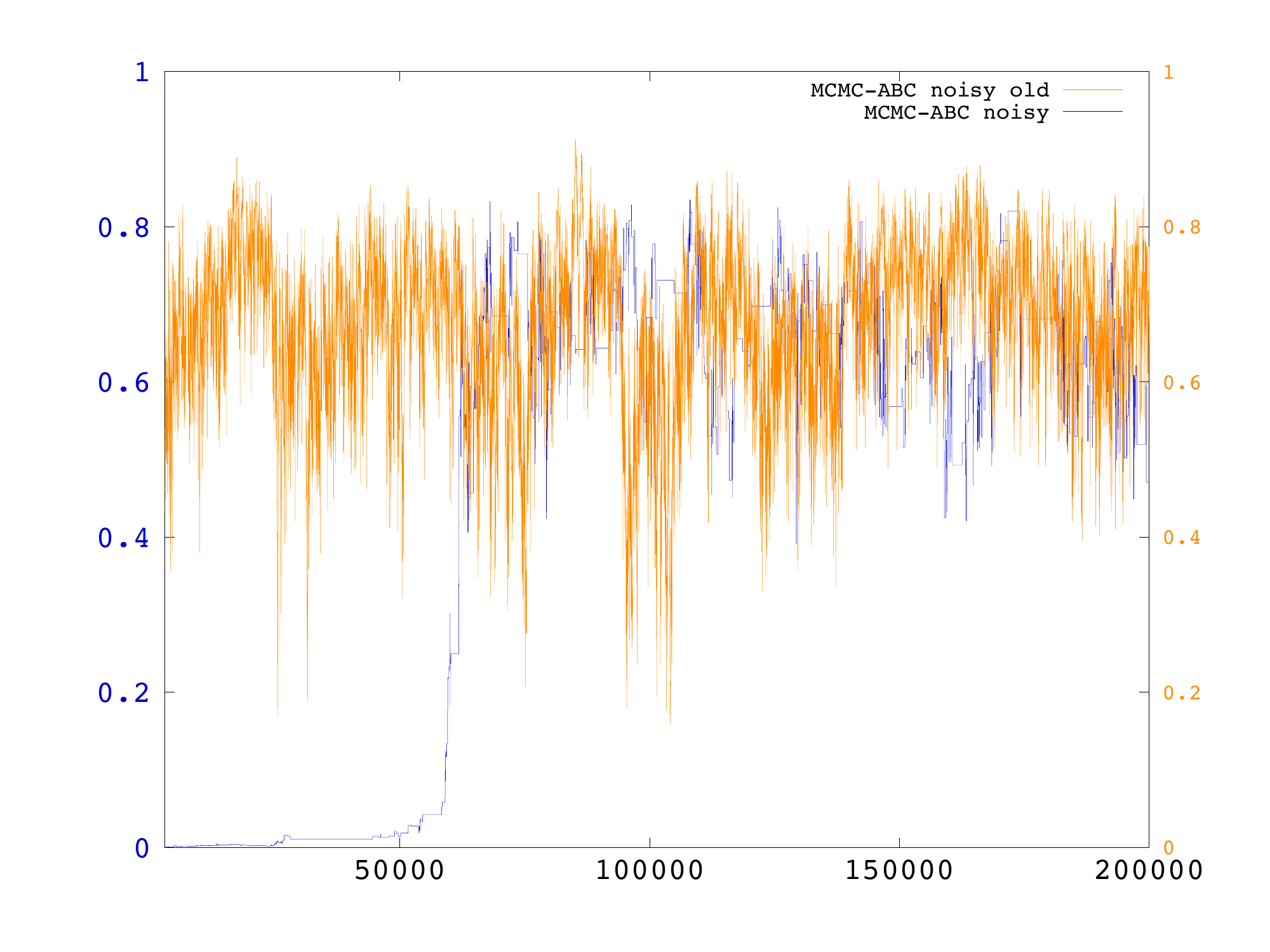}}}
\subfigure[$\beta_2$]{{\includegraphics[width=0.49\textwidth,height=3cm]{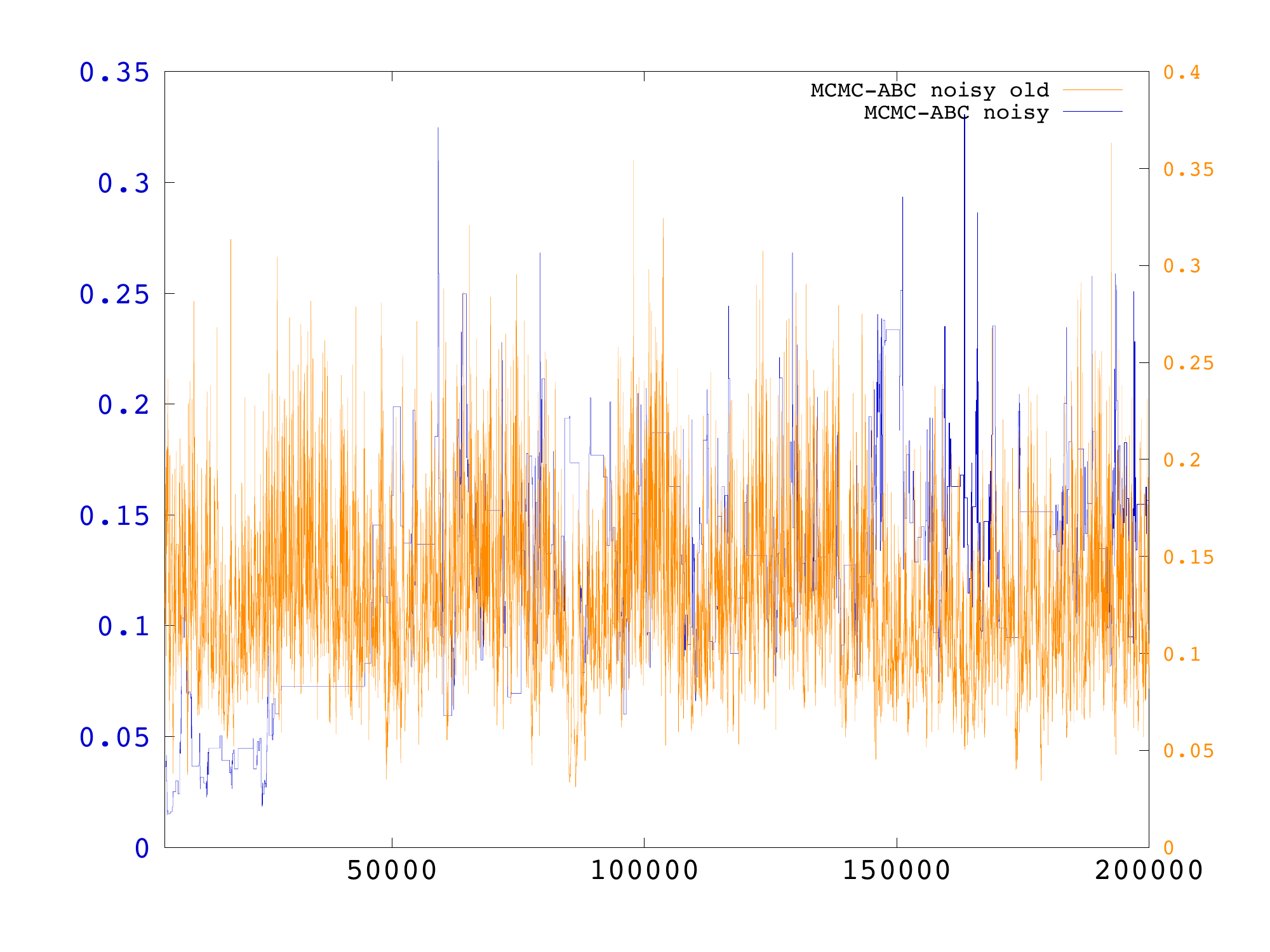}}}
\caption{Trace, $\epsilon=0.01$. We run both algorithms for 200000 iterations, the new kernel is in orange trace and $N=250$ in both cases and the algorithms
are run the S \& P 500 data.}
\label{fig:tracee001} 
\end{figure}

In Figure \ref{fig:tracee001} we can observe the trace plots from a particular (typical) run when $\epsilon=0.01$. In this case, both algorithms are run for 200000 iterations.
Algorithm \ref{alg:Ntry} took about 0.28 seconds per iteration and Algorithm \ref{alg:new} took about 2.06 seconds per iteration; this issue is discussed below.
In this scenario, considerable effort was expended for Algorithm \ref{alg:Ntry} to yield an acceptance rate around
0.3, but despite this, we were unable to make the algorithm traverse the state-space. In contrast, with less effort, Algorithm \ref{alg:new} appears to perform quite well and move around the parameter space (the acceptance rate was around 0.15 versus 0.01 for Algorithm \ref{alg:Ntry}). 
Whilst the computational time for Algorithm \ref{alg:new} is considerably more than Algorithm \ref{alg:Ntry}, in the same amount of computation time, it still moves more around the state
space; algorithm runs of the same length are provided for presentational purposes.
We remark that, whilst we do not claim that it is `impossible' to make Algorithm \ref{alg:Ntry} mix well in this example, we were unable to do so and, alternatively, for Algorithm \ref{alg:new} we expended considerably less effort for very reasonable performance.
This example is typical of many runs of the algorithm and examples we have investigated and is consistent with the discussion in Section \ref{sec:comp_consider}, where we stated
that Algorithm \ref{alg:new} is likely to out-perform Algorithm \ref{alg:Ntry} when the $\alpha_{k}(y_{1:k},\epsilon,\gamma)$
are not large, which is exactly the scenario in this example.

Turning to the cost of simulating Algorithm \ref{alg:new}; for the case $\epsilon=0.5$ we simulated the data an average of 148000 times (per-iteration) and for $\epsilon=0.01$ this figure was 330000. In this example signifcant effort is expended in simulating the $m_{1:n}$.
This shows, at least in this example, that one can run the algorithm without it failing to sample the $m_{1:n}$. The results here suggest that one
should prefer Algorithm \ref{alg:new} only in challenging scenarios, as it can be very expensive in practice.

Finally, we remark that the MLE for a Gaussian Garch model, is $\beta_{0:2} = (4.1\times 10^{-6}, 0.16,0.82)$. This differs to the posterior means, which may indicate that a stable distribution could be useful for modelling the observations for this class of models.

\section{Conclusions}

\label{sec:summ}

In this article we have considered approximate Bayesian inference
from observation driven time series models. We looked at some consistency
properties of the corresponding MAP estimators and also proposed an
efficient ABC-MCMC algorithm to sample from these approximate posteriors.
The performance of the latter was illustrated using numerical examples.

There are several interesting extensions to this work: 
\begin{itemize}
\item  the asymptotic analysis of the ABC posterior in Section \ref{sec:consis}
can be further extended. For example, one may consider Bayesian consistency
or Bernstein Von-Mises theorems, which could provide
further justification to the approximation that was introduced here.
Alternatively, one could look at the the asymptotic bias of the ABC
posterior w.r.t.~$\epsilon$ or the asymptotic loss in efficiency
of the noisy ABC posterior w.r.t.~$\epsilon$ similar to the work
in \cite{dean} for hidden Markov models. 
\item the geometric ergodicity of the presented MCMC sampler can be further
investigated in the spirit of \cite{andrieu1,lee1}. 
\item an investigation to extend the ideas here for sequential Monte Carlo
methods should be beneficial. This has been initiated in \cite{jasra1}
in the context of particle filtering for a different class of models.

\end{itemize}

\section*{Acknowledgements}

A. Jasra acknowledges support from the MOE Singapore and funding from
Imperial College London. N. Kantas was kindly funded by EPSRC under
grant EP/J01365X/1.

\appendix

\section{Proofs for Section \ref{sec:model}}

\begin{proof}{[}Proof of Proposition \ref{prop1}{]} The proof of
$\lim_{n\rightarrow\infty}\mathsf{d}(\theta_{n,x,\epsilon},\Theta_{\epsilon}^{*})=0\quad\mathbb{P}_{\theta*}-a.s.$
follows from \cite[Theorem 21]{douc} if we can establish conditions
(B1-3) for our perturbed ABC model. Clearly (B1) and part of (B2)
holds. (B3-i) hold via \cite[Lemma 21]{douc} via (A\ref{as:3}).
We first need to show that for any $y\in\mathsf{Y}$ that $x\mapsto h^{\epsilon}(x,y)$
is continuous. Consider 
\[
|h^{\epsilon}(x,y)-h^{\epsilon}(x',y)|=\frac{1}{\mu(B_{\epsilon}(0))}|\int_{B_{\epsilon}(y)}\left(h(x,y)-h(x',y)\right)\mu(dy)|.
\]
 Let $\varepsilon>0$, then, by (A\ref{as:2}) there exists a $\delta>0$
such that for $d(x,x')<\delta$ 
\[
\sup_{y\in\mathsf{Y}}|h(x,y)-h(x',y)|<\varepsilon
\]
 and hence for $(x,x')$ as above 
\[
|h^{\epsilon}(x,y)-h^{\epsilon}(x',y)|<\varepsilon.
\]
 which establishes (B2) of \cite{douc}. Now, for (B3-ii) of \cite{douc},
we note that as $\underline{h}\leq h^{\epsilon}(x,y)\leq\overline{h}<\infty$
(see (A\ref{as:2})) the $\log$ function is Lipshitz and 
\[
|\log(h^{\epsilon}(\Phi^{\theta}(Y_{1:k-1})(x),Y_{k}))-\log(h^{\epsilon}(\Phi^{\theta}(Y_{-\infty:k-1})(x),Y_{k}))|\leq C|h^{\epsilon}(\Phi^{\theta}(Y_{1:k-1})(x),Y_{k})-h(\Phi^{\theta}(Y_{-\infty:k-1})(x),Y_{k})|
\]
 for some $C<\infty$ that does not depend upon $Y_{-\infty:k-1},Y_{k},x,\epsilon$.
Now 
\[
|h^{\epsilon}(\Phi^{\theta}(Y_{1:k-1})(x),Y_{k})-h(\Phi^{\theta}(Y_{-\infty:k-1})(x),Y_{k})|=(\mu(B_{\epsilon}(0)))^{-1}|\int_{B_{\epsilon}(Y_{k})}[h(\Phi^{\theta}(Y_{1:k-1})(x),y)-h(\Phi^{\theta}(Y_{-\infty:k-1})(x),y)]\mu(dy)|
\]
 and 
\[
|\int_{B_{\epsilon}(Y_{k})}\left(h(\Phi^{\theta}(Y_{1:k-1})(x),y)-h(\Phi^{\theta}(Y_{-\infty:k-1})(x),y)\right)\mu(dy)|\leq\mu(B_{\epsilon}(0))\sup_{y\in\mathsf{Y}}|h(\Phi^{\theta}(Y_{1:k-1})(x),y)-h(\Phi^{\theta}(Y_{-\infty:k-1})(x),y)]|.
\]
 Thus, by (A\ref{as:3}) and the fact that (B3-i) of \cite{douc}
holds: 
\[
\lim_{k\rightarrow}\sup_{\theta\in\Theta}|\log(h^{\epsilon}(\Phi^{\theta}(Y_{1:k-1})(x),Y_{k}))-\log(h^{\epsilon}(\Phi^{\theta}(Y_{-\infty:k-1})(x),Y_{k}))|=0\quad\mathbb{P}_{\theta^{*}}-a.s.
\]
 Note, finally that (B3-iii) trivially follows by $h^{\epsilon}(x,y)\leq\overline{h}<\infty$.
Hence we have proved that 
\[
\lim_{n\rightarrow\infty}\mathsf{d}(\theta_{n,x,\epsilon},\Theta_{\epsilon}^{*})=0\quad\mathbb{P}_{\theta*}-a.s..
\]
 \end{proof}

\begin{proof}{[}Proof of Proposition \ref{prop2}{]} This result
follows from \cite[Proposition 23]{douc}. One can establish assumptions
(B1-3) of \cite{douc} using the proof of Proposition \ref{prop1}.
Thus we need only prove that 
\[
H^{\epsilon}(x,\cdot)=H^{\epsilon}(x',\cdot),\quad\Leftrightarrow x=x'.
\]
 Now, for any $A\in\mathcal{Y}$ 
\[
H^{\epsilon}(x,A)=\frac{1}{\mu(B_{\epsilon}(0))}\int_{A}[\int_{B_{\epsilon}(y)}H(x,du)]\mu(dy).
\]
 By (A\ref{as:4}) $\int_{B_{\epsilon}(y)}H(x,du)=\int_{B_{\epsilon}(y)}H(x',du)$
$\Leftrightarrow x=x'$, so 
\[
H^{\epsilon}(x,A)=\frac{1}{\mu(B_{\epsilon}(0))}\int_{A}[\int_{B_{\epsilon}(y)}H(x',du)]\mu(dy)\Leftrightarrow x=x'
\]
 which completes the proof. \end{proof}

\section{Proof for Section \ref{sec:comp}}

\begin{proof}{[}Proof of Proposition \ref{prop:rel_var}{]} We have
\[
\mathbb{E}_{\gamma,N}\bigg[\bigg(\frac{\prod_{k=1}^{n}\frac{1}{M_{k}-1}}{\prod_{k=1}^{n}\frac{\alpha_{k}(y_{1:k},\epsilon,\gamma)}{N-1}}-1\bigg)^{2}\bigg]=\frac{1}{(\prod_{k=1}^{n}\frac{\alpha_{k}(y_{1:k},\epsilon,\gamma)}{N-1})^{2}}\bigg(\prod_{k=1}^{n}\mathbb{E}_{\gamma,N}\Big[\frac{1}{(M_{k}-1)^{2}}\Big]-\Big(\prod_{k=1}^{n}\frac{\alpha_{k}(y_{1:k},\epsilon,\gamma)}{N-1}\Big)^{2}\bigg).
\]
 Now, by \cite{neuts,zacks} ($N\geq3$) for any $k\geq1$ 
\[
\mathbb{E}_{\gamma,N}\Big[\frac{1}{(M_{k}-1)(M_{k}-2)}\Big]=\frac{\alpha_{k}(y_{1:k},\epsilon,\gamma)^{2}}{(N-1)(N-2)}
\]
 and thus clearly 
\[
\mathbb{E}_{\gamma,N}\Big[\frac{1}{(M_{k}-1)^{2}}\Big]\leq\frac{\alpha_{k}(y_{1:k},\epsilon,\gamma)^{2}}{(N-1)(N-2)}.
\]
 hence 
\begin{equation}
\mathbb{E}_{\gamma,N}\bigg[\bigg(\frac{\prod_{k=1}^{n}\frac{1}{M_{k}-1}}{\prod_{k=1}^{n}\frac{\alpha_{k}(y_{1:k},\epsilon,\gamma)}{N-1}}-1\bigg)^{2}\bigg]\leq(N-1)^{2n}\Big(\frac{1}{(N-1)^{n}(N-2)^{n}}-\frac{1}{(N-1)^{2n}}\Big).\label{eq:maineq1}
\end{equation}
 Now the R.H.S.of \eqref{eq:maineq1} is equal to 
\begin{equation}
\frac{nN^{n-1}+\sum_{i=2}^{n}\binom{n}{i}N^{n-i}[(-1)^{i}-(-2)^{i}]}{N^{n}-2nN^{n-1}+\sum_{i=2}^{n}\binom{n}{i}N^{n-i}(-2)^{i}}.\label{eq:maineq2}
\end{equation}
 Now, we will show 
\begin{equation}
\sum_{i=2}^{n}\binom{n}{i}N^{n-i}[(-1)^{i}-(-2)^{i}]\leq0.\label{eq:ineq1}
\end{equation}
 The proof is given when $n$ is odd. The case $n$ even follows by
the proof as $n-1$ is odd and the additional term is negative. Now
we have for $k\in\{1,3,\dots,(n-1)/2\}$ that the sum of consecutive
even and odd terms is equal to 
\[
\frac{N^{n-2k}n!}{(n-2k-1)!(2k)!}\bigg[\frac{N(1-2^{2k})(2k+1)-(2^{2k+1}-1)(n-2k)}{(n-2k)(2k+1)N}\bigg]
\]
 which is negative as 
\[
N\geq\frac{(2^{2k+1}-1)(n-2k)}{(1-2^{2k})(2k+1)}.
\]
 Thus we have established \eqref{eq:ineq1}. We will now show that
\begin{equation}
\sum_{i=2}^{n}\binom{n}{i}N^{n-i}(-2)^{i}\geq0.\label{eq:ineq2}
\end{equation}
 Following the same approach as above (i.e.~$n$ is odd) the sum
of consecutive even and odd terms is equal to 
\[
\frac{N^{n-2k}2^{2k}n!}{(n-2k-1)!(2k)!}\bigg[\frac{N(2k+1)-2(n-2k)}{(n-2k)(2k+1)N}\bigg].
\]
 This is positive if 
\[
N\geq\frac{n-2k}{2k+1}\leq\frac{n}{3}.
\]
 as $N\geq\frac{2n}{(1-\beta)}$ and $6\geq(1-\beta)$ it follows
that $N\geq n/3\geq(n-2k)/(2k+1)$; thus one can establish \eqref{eq:ineq2}.

Now returning to \eqref{eq:maineq1} and noting \eqref{eq:maineq2},
\eqref{eq:ineq1} and \eqref{eq:ineq2}, we have 
\[
\mathbb{E}_{\gamma,N}\bigg[\bigg(\frac{\prod_{k=1}^{n}\frac{1}{M_{k}-1}}{\prod_{k=1}^{n}\frac{\alpha_{k}(y_{1:k},\epsilon,\gamma)}{N-1}}-1\bigg)^{2}\bigg]\leq\frac{nN^{n-1}}{N^{n}-2nN}=\frac{n}{N-2n}
\]
 as $N\geq2/(1-\beta)$ it follows that $n/(N-2n)\leq Cn/N$ and we
conclude.

\end{proof}

\begin{proof}{[}Proof of Proposition \ref{prop:time}{]} We have
\begin{eqnarray*}
\mathbb{E}_{\zeta K^{i}\otimes\tilde{Q}}[\sum_{k=1}^{n}M_{k}] & = & \int_{(\Theta\times\mathsf{X})^{2}}\sum_{\mathsf{M}_{N}^{n}}\Big(\sum_{k=1}^{n}m_{k}\Big)\Big\{\prod_{k=1}^{n}\binom{m_{k}-1}{N-1}\alpha_{k}(y_{1:k},\gamma',\epsilon)^{N}(1-\alpha_{k}(y_{1:k},\gamma',\epsilon))^{m_{k}-N}\Big\} q(\gamma,\gamma')\zeta K^{i}(d\gamma)d\gamma'\\
 & = & \int_{(\Theta\times\mathsf{X})^{2}}\Big(\sum_{k=1}^{n}\frac{N}{\alpha_{k}(y_{1:k},\gamma,\epsilon)}\Big)q(\gamma,\gamma')\zeta K^{i}(d\gamma)d\gamma'\leq\frac{nN}{C}.
\end{eqnarray*}
 where we have used the expectation of a negative-binomial random
variable and applied $\inf_{k}\alpha_{k}(y_{1:k},\gamma,\epsilon)\geq C$,
$\mu-$a.e. in the inequality 
\end{proof}

\end{document}